\documentclass[journal]{IEEEtran}

\ifCLASSINFOpdf
   \usepackage[pdftex]{graphicx}
   \graphicspath{{../pdf/}{../jpeg/}}
   \DeclareGraphicsExtensions{.pdf,.jpeg,.png}
\else
   \usepackage[dvips]{graphicx}
   \graphicspath{{../eps/}}
   \DeclareGraphicsExtensions{.eps}
\fi

\usepackage{cuted}
\usepackage{lipsum, color}

\usepackage[noadjust]{cite}
\usepackage{filecontents}

\usepackage{subfig}

\usepackage{bm}

\newcommand{\specialcell}[2][c]{
  \begin{tabular}[#1]{@{}c@{}}#2\end{tabular}}

\usepackage{amsmath}
\usepackage{amssymb}
\usepackage{algorithm}
\usepackage{algorithmicx}
\usepackage[noend]{algpseudocode}
\newcommand{\argmax}{\operatornamewithlimits{argmax}}
\newcommand{\argmin}{\operatornamewithlimits{argmin}}
\usepackage{epstopdf}

\newcommand{\rom}[1]{%
  \textup{\uppercase\expandafter{\romannumeral#1}}%
}

\begin{document}
%
\title{Acoustic Reflector Localization: Novel Image Source Reversion and Direct Localization Methods}
%
%
%

\author{Luca~Remaggi,~\IEEEmembership{Student Member~IEEE},
        Philip~J.~B.~Jackson,
        Philip Coleman, \\
        Wenwu Wang,~\IEEEmembership{Senior~Member~IEEE}
\thanks{Color versions of one or more of the figures in this paper are available online
at http://ieeexplore.ieee.org. 
Digital Object Identifier 10.1109/TASLP.2016.2633802. Manuscript received May 13, 2016; revised September 23, 2016 and November 14, 2016; accepted November 22, 2016. Date of publication December 1,
2016. This work was supported by the Engineering and Physical Sciences Research Council (EPSRC) Grants EP/K014307/1 and EP/L000539/1, and the MOD University Defence Research Collaboration in Signal Processing. The authors are with the Centre for Vision, Speech and Signal Processing, University of Surrey, Guildford, GU2 7XH, UK, emails: [l.remaggi,~p.jackson,~p.d.coleman,~w.wang]@surrey.ac.uk.}}

\markboth{IEEE/ACM TRANSACTIONS ON AUDIO, SPEECH, AND LANGUAGE PROCESSING,~Vol.~25, No.~2, February 2017}{Remaggi \MakeLowercase{\textit{et al.}}: Acoustic Reflector Localization: Image-Source Reversion or TDOA-based Reconstruction?} 

\maketitle

\begin{abstract}
Acoustic reflector localization is an important issue in audio signal processing, with direct applications in spatial audio, scene reconstruction, and source separation. Several methods have recently been proposed to estimate the 3D positions of acoustic reflectors given room impulse responses (RIRs). In this article, we categorize these methods as ``image-source reversion'', which localizes the image source before finding the reflector position, and ``direct localization'', which localizes the reflector without intermediate steps. We present five new contributions. First, an onset detector, called the clustered dynamic programming projected phase-slope algorithm, is proposed to automatically extract the time of arrival for early reflections within the RIRs of a compact microphone array. 
Second, we propose an image-source reversion method that uses the RIRs from a single loudspeaker. It is constructed by combining an image source locator (the image source direction and range (ISDAR) algorithm), and a reflector locator (using the loudspeaker-image bisection (LIB) algorithm). Third, two variants of it, exploiting multiple loudspeakers, are proposed. Fourth, we present a direct localization method, the ellipsoid tangent sample consensus (ETSAC), exploiting ellipsoid properties to localize the reflector. Finally, systematic experiments on simulated and measured RIRs are presented, comparing the proposed methods with the state-of-the-art. ETSAC generates errors lower than the alternative methods compared through our datasets. Nevertheless, the ISDAR-LIB combination performs well and has a run time 200 times faster than ETSAC. 
\end{abstract}


\IEEEpeerreviewmaketitle

\section{Introduction}
\label{sec:introduction}
The creation of an accurate model to identify acoustic reflector positions from room impulse responses (RIRs) is important for several different research areas in audio signal processing. For instance, such a model can provide information about the geometry of the listening environment with respect to a listening position, which can be exploited in audio forensics~\cite{Malik2013}, simultaneous localization and mapping \cite{DokDauVet2016}, and spatial audio \cite{ZotDurDav2004}. In addition, model parameters can be utilized to enhance target signals in fields such as automatic speech recognition \cite{YosSehDelKinMaaNakKel2012}, music transcription \cite{PluAbdBelDavMonSan2002}, source separation \cite{AsaGolBouCev2014}, audio tracking \cite{OcaDokVet2014}, dereverberation \cite{NayGau2010}, and microphone array raking, that captures and combines individual reflections constructively~\cite{DokSchVet2015, JavMooNay2016}. Beyond all of these, an acoustic reflection localizer can be combined with image processing to construct a robust hybrid room geometry model \cite{HussCivMon2014,YeZhaYanMan2015}. In particular, acoustic information can aid detection of mirrors or windows, that cannot be identified by a visual sensor. 
   
Several approaches have been presented to localize reflectors from RIRs~\cite{CancAnnAntSarRabTub2011,CanAntThoFilSarNayTub2011,DokLuVett2011,ArtGarMatUsh2013,MarAntSarTub2013,MarAntSarTub2013_IEEE,MooreBrookesNaylor,AntFilThomHabSarNayTub}, as the solution of a 2D problem, where the loudspeaker, microphone and reflector lie on the same plane. For instance, in \cite{AntFilThomHabSarNayTub}, the authors exploited time of arrival (TOA) knowledge to localize 2D reflectors, using ellipses. These ellipses have their major axis equal to a specific reflection path, and foci on the respective microphone and source positions, and the common tangent line corresponds to the 2D reflector under investigation. Recently, 3D models have also been investigated. In \cite{FilCanAntSarNay2012}, the work in \cite{AntFilThomHabSarNayTub} was extended to 3D spaces. However, it was not yet a full 3D estimation, instead combining 2D projections to estimate the positions of surfaces outside the measurement planes. A different way to extend the work in \cite{AntFilThomHabSarNayTub} was proposed in \cite{RemJackColWan2014}, where we considered the ellipses generated to be 2D projections of ellipsoids. This improved the accuracy compared to \cite{AntFilThomHabSarNayTub}, but was also not fully 3D.
The 3D reflector localization methods can approximately be categorized into two groups. The first one is ``image-source reversion''~\cite{RibFloBaZha2012, TervoTossa2012, DokParWalLuVet2013}, where the TOA is used to revert to the image source location \cite{AllenBerkley79}, which is subsequently used to determine the reflector position. The second group contains those methods directly localizing the reflector, without estimating other elements about the room acoustic first~\cite{KusdeVHulGis2004, ZamanAnnRab2014, NasAntSarTub2011}. Accordingly, we refer to this group as ``direct localization''. 

The method in \cite{TervoTossa2012} used the image-source reversion approach to localize reflectors in 3D. In \cite{TervoTossa2012}, a large number of loudspeaker orientations were needed. In \cite{RibFloBaZha2012}, a least-squares minimization was utilized to fit ``synthetic'' reflections to recorded RIRs. However, it still required a large number of RIRs. The synthetic reflections were obtained in an anechoic room, with a plastic reflector to simulate a wall. In total, 240 loudspeaker positions were used to collect the recordings, leading to 1440 RIRs, considering the six-element microphone array employed. 
Since the number of RIRs was deemed not enough to train the model, the RIRs were additionally interpolated in space, in order to have more directions of arrival (DOAs). The main contribution of \cite{DokParWalLuVet2013} was a novel algorithm to label the reflections from a distributed microphone array, where the reflector order would otherwise be ambiguous if compared among different microphone recordings. The reflector estimation was performed using image-source reversion, by assuming that the TOAs were known a priori.  

In \cite{RemJackCol2015AES}, we proposed the image source direction and ranging (ISDAR) method. ISDAR was based on the idea of combining TOA and DOA to localize image sources with the following four novel aspects. The first aspect is the use of a compact array of non-coincident omnidirectional microphones. This gives a marked improvement with respect to \cite{Gunel2002}, where recordings were made with a coincident first-order directional microphone (B-format). The second aspect concerns the TOA estimation. In our approach, TOAs are estimated for each microphone channel by the dynamic programming projected phase-slope algorithm (DYPSA) \cite{NayKouGudBroo}, and then clustered. This allows for the development of more robust algorithms, by detection and correction of gross errors. In \cite{Gunel2002}, a virtual cardioid is used to scan and find the DOA and TOA of the reflection, at the maximum amplitude of the directional signal. Correlations between each pair of microphone RIRs were utilized in \cite{TerPatKuuLokk2013}, where a squared-error cost function was then minimized to find DOAs from estimated TDOAs. Third, in \cite{Gunel2002}, the DOA from the reflection was taken as that whose directional response was maximum. Later, in  \cite{TerPol2015}, a probabilistic approach was proposed to find DOA by steering towards the signals recorded through spherical microphone arrays. In our case the delay-and-sum beamformer, which is a TDOA-based approach, was employed to obtain the DOA as that giving the maximum response. Finally, whereas direct sound cancellation can avoid swamping the reflection signal \cite{RobXia2010}, for ISDAR we chose to apply a time window around the reflection TOA, to gate the reflection signal and extract the related time domain segment, as in \cite{Gunel2002, TerPol2015}.

One of the first attempts to employ, instead, the direct localization approach was proposed in \cite{KusdeVHulGis2004}. Exploiting inverse wave field extrapolation, the authors mapped reflections from a set of receivers to the related reflecting objects. This method provided an accurate analysis, and it was even able to identify small reflecting objects lying between the main reflector and the microphone array. However, the microphone array was assumed to be exactly parallel to the reflector. In \cite{ZamanAnnRab2014}, reflector positions were estimated by transforming the RIR into the frequency domain. In spite of this, only two parallel reflectors were localized, assuming the other boundaries completely absorbent.
The method that can be considered as the first one exploiting direct localization through 3D ellipsoids, was presented in \cite{NasAntSarTub2011}. Despite this, the peaks were not automatically identified to extract TOAs from RIRs. In addition, the reflector search was computationally expensive, caused mainly by the optimization of its cost function. In our previous work \cite{RemJackWangChamb2015}, a full 3D method was proposed, exploiting direct reconstruction through ellipsoids. The reflector search was performed utilizing two variants, either the random sample consensus (RANSAC) algorithm~\cite{Fischler1981}, or the combination of the cost function used in \cite{NasAntSarTub2011} and the Hough transform.  

Methods to localize sources that combine TOA with time difference of arrival (TDOA) can be found in the literature. For instance, in \cite{TerPatLok2012}, the authors provided an evaluation over several localization methods, exploiting the TOA and TDOA probability density functions. Then, they fused together these two densities, improving robustness. However, with ISDAR, we combine TOA and DOA directly, in spherical coordinates. 

We have observed various limitations in the reflector localization literature. There is not a onset detection algorithm specifically designed to automatically extract reflection TOAs from multichannel RIRs, recorded by compact microphone arrays; although several algorithms have been designed for single channel onset detection on RIRs~\cite{VOICEBOX1997, Kuster2008, DefDauPol2009, Usher2010}, including one based on spatial pre-processing of B-Format signals~\cite{Gunel2002}. Some methods assume specific spatial relationships between the microphone array and reflectors~\cite{KusdeVHulGis2004, ZamanAnnRab2014}. Methods may require thousands of RIRs recorded in anechoic rooms~\cite{RibFloBaZha2012}. The microphones are often considered to be spatially sparse, introducing the problem of labeling echoes \cite{DokParWalLuVet2013}. There are no proposed ways to aggregate measurements from multiple loudspeakers to improve localization. Finally, classical image-source reversion methods (e.g. \cite{TervoTossa2012, DokParWalLuVet2013}) use TOAs to localize the image source without considering other information carried by the RIRs, limiting their robustness. 

To address these issues, the contributions of this work are: 
\begin{itemize}
\item a multichannel version of DYPSA \cite{NayKouGudBroo}, i.e. clustered DYPSA (C-DYPSA), to automatically extract reflection TOAs from compact microphone array RIRs;
\item the image-source reversion method ISDAR-LIB, created by the fusion of our ISDAR (the image-source locator presented in \cite{RemJackCol2015AES}) and loudspeaker-image bisection (LIB) (a reflector localization algorithm in \cite{TervoTossa2012, DokParWalLuVet2013});
\item two further novel variants of ISDAR-LIB, exploiting multiple loudspeakers;
\item ellipsoid tangent sample consensus (ETSAC), a direct localization method (modified from \cite{RemJackWangChamb2015}, by utilizing the new C-DYPSA instead of DYPSA);
\item a comparative evaluation of the state-of-the-art and the proposed methods, using synthetic and measured RIRs.
\end{itemize}
The comprehensive comparison presented here is, to our knowledge, the first that compares image-source reversion and direct localization methods, as approaches for 3D reflector localization. The study also informs the level of estimation accuracy expected from a real-world dataset~\cite{RemJacColFra2015}.

The rest of the paper is organized as follows: Section \ref{sec:background_theory} introduces the underlying theory supporting the presented methods, and the pre-processors common to every method evaluated. The state-of-the-art methods selected for the evaluation are described in Section \ref{sec:State-of-the-Art Methods}, and the proposed methods in Section \ref{sec:proposes_methods}. The numerical analysis and results are reported in Section \ref{sec:exp_eval}. Section \ref{sec:concl} draws the overall conclusions.

\section{Background and Preliminaries}
\label{sec:background_theory}

\subsection{Theoretical Models}

\subsubsection{The Room Impulse Response (RIR)}
\label{subsec:RIR}
A RIR is an acoustic signal, carrying information about the environment in which it was recorded. It is generally considered as being composed of three components \cite{Kuttruff4}: the direct sound, revealing the position of the sound source; the early reflections, conveying a sense of the environmental geometry; and the late diffuse reverberation, indicating the size of the environment, and average absorption~\cite{Blesser2001, ValParSavSmiAbe2012}. From this classical decomposition, and defining the discrete time variable $n$, the RIR from source $j$ to sensor $i$ can be defined as superimposition of Dirac deltas, delayed by $n_{k,i,j}$ samples, with $k$ enumerating the reflections~\cite{RemJackCol2015AES}:
\begin{equation}
\begin{aligned}
r_{i,j}(n)=&h_{0,i,j}(n-n_{0,i,j}) + \sum\limits_{k=1}^{T_m}h_{k,i,j}(n-n_{k,i,j}) + g(n),
\end{aligned}
\label{eq:RIR_real}
\end{equation}
where $h_{0,i,j}(n)$ represents the direct sound, $h_{k,i,j}(n)$ the early reflections, and $g(n)$ is the late reverberation modeled as exponentially decaying Gaussian noise; $T_m$ is the $k$-th peak corresponding to the last reflection before the mixing time.

\subsubsection{The Image Source Model}
\label{subsec:image_source}
The most prominent early reflections typically have a sizable specular component. Therefore, one way to localize reflectors is to benefit from the notion of image sources \cite{AllenBerkley79}. Denoting $\textbf{B}_0$ as a vector containing the three Cartesian coordinates of the sound source and $\textbf{p}$ as a vector containing the first reflector ($k=1$) plane coefficients, the corresponding image source $\textbf{B}$ is the virtual sound source, constructed as the point $\textbf{B}_0$ mirrored with respect to $\textbf{p}$. 

\begin{table}[!t]

\caption{3D method categorization.}
\label{tab:3d_models}
\centering

\begin{tabular}{|c|c|}
\hline
\textbf{Image-Source Reversion} & \textbf{Direct Localization} \\
\hline
Ribeiro \emph{et al.} \cite{RibFloBaZha2012} & Kuster \emph{et al.} \cite{KusdeVHulGis2004}\\
\hline
Tervo \emph{et al.} \cite{TervoTossa2012} & Zamaninezhad \emph{et al.} \cite{ZamanAnnRab2014} \\ 
\hline
Dokmani\'{c} \emph{et al.} \cite{DokParWalLuVet2013} & Nastasia \emph{et al.} \cite{NasAntSarTub2011} \\
\hline
Proposed ISDAR-LIB and variants & Proposed ETSAC \\
\hline
\end{tabular}
\end{table}

\subsection{Method Classification and Overviews}
\label{subsec:overall_flows}
\begin{figure}[t]
\centering
\includegraphics[width=\columnwidth]{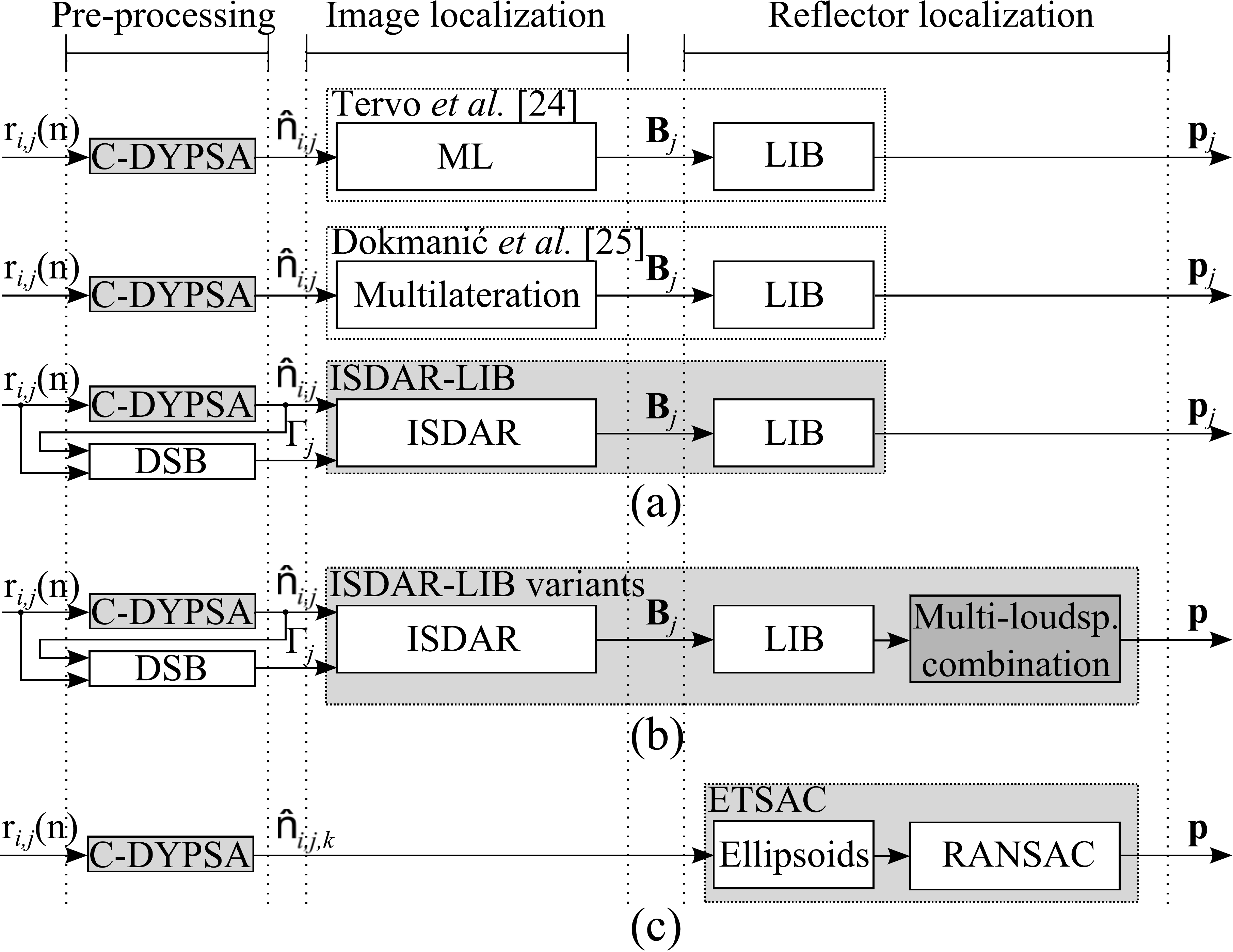}
\caption{Overview of the tested methods (image-source reversion methods exploiting a single loudspeaker (a), image-source reversion methods exploiting multiple loudspeakers (b), direct localization method (c)). The novel methods are highlighted in grey. $r_{i,j}(n)$ is the RIR between the $i$-th microphone and $j$-th loudspeaker; $\hat{\textbf{n}}_{i,j,k}$ is the TOA of the direct sound ($k=0$), or first reflection ($k=1$); $\Gamma_{j}=[\Theta_{j},\Phi_{j}]$ contains the azimuth $\Theta_{j}$ and elevation $\Phi_{j}$; $\textbf{B}_j$ is the image source position; $\textbf{p}_j$ is the plane related to the $j$-th loudspeaker, whereas $\textbf{p}$ is obtained using multiple loudspeakers.}
\label{fig:all_models_overview}
\end{figure}

As discussed in the introduction, reflector locator methods can be divided into two main groups: image-source reversion and direct localization. Table \ref{tab:3d_models} summarizes these groups and shows the categorization for our proposed methods, together with the state-of-the-art.
Figure \ref{fig:all_models_overview} shows an overview of the structure of the proposed methods, together with two baseline methods (selected to be compared in our experiments in Section \ref{sec:exp_eval}): Tervo \emph{et al.} \cite{TervoTossa2012} and Dokmani\'{c} \emph{et al.} \cite{DokParWalLuVet2013}. The novel methods and algorithms are highlighted in grey (C-DYPSA, ISDAR-LIB, the ISDAR-LIB variants, and ETSAC). To generate methods that are able to automatically extract TOAs from RIRs, C-DYPSA is proposed and employed as a pre-processor to each method tested. A description of this novel algorithm, that is an evolution of the DYPSA algorithms \cite{NayKouGudBroo}, is reported in Section \ref{subsubsec:C-DYPSA}. Different acoustic parameters are exploited by the methods. 
ISDAR-LIB and its variants exploit DOA to localize the reflector. Thus, for these methods, the delay-and-sum beamformer~(DSB)~\cite{VanVeenBuck1988} is employed in the pre-processing stage, together with C-DYPSA.

\subsection{Method Assumptions}
\label{subsec:assumptions}
The main assumptions in this article are as follows:
\begin{itemize}
\item knowledge of at least four RIRs;
\item the omnidirectional microphone array is ``compact'';
\item the sources are in the far-field;
\item the reflection has a dominant specular component;
\item the image source estimation errors are independent and identically distributed.
\end{itemize}
The assumption concerning the minimum number of RIRs has been made due to the fact that, to estimate parameters in 3D space, at least four positions are needed. The “compact” microphone array assumption has been made to enable the use of classical beamforming techniques \cite{VanVeenBuck1988}, and to avoid erroneous permutations in the labelling of reflections arriving at the microphone \cite{DokParWalLuVet2013}. We consider arrays with maximum microphone displacement from the array center less than half a 1-kHz wavelength ($d<171$\,mm) to be compact, where 1\,kHz is a standard reference frequency. Note that compactly-arranged microphones are similarly affected by any source directivity. Assuming sources and image sources to be in the far-field means their response at the array can be approximated as plane waves. For the array configuration in the present work, the Fraunhofer rule sets the far-field limit at 2.1\,m at 1\,kHz \cite{Balanis}, making this a fair assumption for reflector localization in typical rooms. The fourth assumption, of the specular reflections, enables its detection in the time domain RIR. This assumption excludes scattering, shadowing and diffraction phenomena, and justifies the use of the image source model, which applies to mid-range audio frequencies. No further assumptions regarding reflection, source, and microphone frequency responses are needed. Finally, assuming the image source localization errors as independent and identically distributed allows the integration of multiple loudspeaker results in a post-processing step. Different reflection signal-to-noise ratios (SNRs) do not influence this, since the dominant specular component of the reflection implies a high SNR.

The first reflection can be considered as the most important one, for two main reasons~\cite{HowAng2009}: subject to masking, it has the most prominent perceptual properties, being a single specular reflection arriving with a limited time delay from the direct sound; it carries most energy. In a typical room, the second reflection conveys 20--40\,dB less energy than the direct sound~\cite{HowAng2009}. Therefore, in this article, we focus on the first reflector. Other works that examine later reflections can be found in~\cite{TervoTossa2012, DokParWalLuVet2013}. Since the first reflection is estimated from multichannel RIRs only, we do not have to make any prior assumptions about the room shape or the reflector orientation.

\subsection{Common Pre-processing}
\label{subsec:pre-processing}
To obtain the TOAs and DOAs automatically from recorded RIRs, a pre-processing stage was employed consisting of a clustering onset detector (i.e. C-DYPSA) and a DSB. 

\subsubsection{The Clustered Dynamic Programming Projected Phase-Slope Algorithm (C-DYPSA)}
\label{subsubsec:C-DYPSA}
The state-of-the-art reflector localization methods do not provide a specific algorithm to extract TOAs from the RIRs. In \cite{RemJackWangChamb2015, RemJackCol2015AES}, we applied DYPSA to a single RIR at a time, to detect variations in the amplitude. Here, we propose an extension, that clusters TOAs across the microphones of the compact array.

DYPSA was designed to estimate glottal closure instances from speech \cite{NayKouGudBroo}. Defining the phase-slope function $S(\omega)$ as the opposite of the group delay function of the signal $S(\omega)=-G(\omega)$, peaks in the time domain (i.e. TOAs) correspond to positive-going zero crossings in $S(\omega)$. To reliably select the instants where $S(\omega)$ has these zero crossings, it is smoothed using a Hann window $w_{\mathrm{GD}}(n)$ of length $T_{GD}$. Finally, two processes are applied: to form a confidence level by comparing $S(\omega)$ to a model, and to calculate the weighted gain of each peak considering its importance on the original signal. To adapt the algorithm to our purpose \cite{RemJackCol2015AES}, a threshold $\tau_S$ was defined on $S(\omega)$ to take only the most significant peaks of $r_{i,j}(n)$. Another threshold $\tau_A$ was applied on the time domain amplitude, to eliminate the peaks that have much less energy than the main one. These thresholds were heuristically derived. 

The proposed C-DYPSA contains two post-processing refinements. First, considering the $k$-th peak for every $i$-th microphone in the compact array, the median $\tilde{n}_{k,j}$ of the estimated TOAs in samples $\hat{n}_{k,i,j}$ is obtained. 
The output of DYPSA, considering each $i$-th sensor separately, is then observed. If the $(k+1)$-th reflection TOA $\hat{n}_{k+1,i,j}$ is closer to the median $\tilde{n}_{k,j}$ than the $k$-th reflection TOA $\hat{n}_{k,i,j}$, then $\hat{n}_{k,i,j}$ is treated as a false positive. Consequently, it is replaced with the $k$-th reflection TOA value $\hat{n}_{k+1,i,j}$. Second, the Grubbs' test \cite{Grubbs1969} identifies the cluster of TOAs relative to the $k$-th peak considering every microphone in the compact array. The RIRs generating outliers to this cluster are discarded. 
The C-DYPSA performance is evaluated in Section \ref{subsec:C-DYPSA_eval}, Table \ref{tab:C-DYPSA_eval}.  

\subsubsection{Delay-and-Sum Beamformer (DSB)}
\label{subsubsec:DSB}
Our image-source reversion methods also exploit DOAs \cite{RemJackCol2015AES}. To extract them directly from recorded RIRs, the DSB \cite{VanVeenBuck1988} was used, providing adequate performance for our purposes, and being simple.
To apply the DSB, the input RIRs were first segmented. To generate these segments without losing the phase differences, the average of the first early reflection TOAs $\hat{n}_{i,j}$ over the $M$ microphones was calculated as $\overline{n}_{j}=\frac{1}{M}\sum_{i=1}^M\hat{n}_{i,j}$.  
This TOA corresponds to that of a virtual microphone lying at the center of the array. The segments were obtained by applying a Hamming window $w(n)$ of length $T$, for each RIR, centered at $\overline{n}_{j}$: $r^{S}_{i,j}(n) = r_{i,j}(n)w(n- \overline{n}_{j})$.

\section{State-of-the-Art Methods}
\label{sec:State-of-the-Art Methods}
The two image-source reversion baselines, based on single loudspeaker information, are here presented~\cite{TervoTossa2012, DokParWalLuVet2013}. First, their two distinct image source locator algorithms are described. Then, their common reflector position estimation algorithm LIB is presented. 

\subsection{Maximum Likelihood (ML)} 
\label{subsubsec:Tervo}
The method proposed by Tervo \emph{et al.} in \cite{TervoTossa2012} is composed by a maximum likelihood (ML) algorithm to localize the image source, followed by the LIB algorithm to estimate the reflector position (Figure \ref{fig:all_models_overview}(a)).

The ML image source locator exploits the TOAs at each microphone to generate a probability function \cite{TervoTossa2012}. First, a uniformly distributed set of $x$ points is generated in 3D space, represented by the $x\times 3$ matrix $\textbf{X}$ containing all the Cartesian coordinates. Considering these points as possible image source positions, and assuming the center of the microphone array as the origin of the coordinate system, the possible TOAs for the first reflection of the $j$-th loudspeaker are obtained, and placed in the vector $\textbf{n}_{j}(\textbf{X})=[n_{1,j}(\textbf{X}),...,n_{M,j}(\textbf{X})]^T$, where $M$ is the number of microphones. With the TOAs estimated through C-DYPSA $\hat{\textbf{n}}_{j}=[\hat{n}_{1,j},...,\hat{n}_{M,j}]^T$, the multivariate Gaussian probability distribution function can be calculated:
\begin{equation}
p(D_{j}(\textbf{X}),\Sigma) = \frac{\exp(-\frac{1}{2} D_{j}(\textbf{X})^T  \Sigma^{-1} D_{j}(\textbf{X}))}{ (2\pi)^{M/2} \sqrt{\det(\Sigma)}},
\label{eq:probability_function}
\end{equation}
where $D_{j}(\textbf{X})=\textbf{n}_{j}(\textbf{X})-\hat{\textbf{n}}_{j}$, and $\Sigma=[\mathrm{diag}(J_{i,j})]^{-1}$. $J_{i,j}$ is the Fisher information, carrying knowledge of the selected frame $r^S_{i,j}(n)$ SNR~\cite{TerPatLok2012}. Thus, the image position is given by: 
\begin{equation}
\textbf{B}_{j}=\argmax_{\textbf{X}} p(D_{j}(\textbf{X}),\Sigma).
\label{eq:image}
\end{equation}

\subsection{Multilateration}
The method presented by Dokmani\'{c} et al. \cite{DokParWalLuVet2013} employs the image localization method named multilateration\footnote{Multilateration is not explicitly presented in \cite{DokParWalLuVet2013}. However, it can be identified from the authors' code at http://infoscience.epfl.ch/record/186657/files/.}. In addition, the LIB algorithm, which will be introduced in Section \ref{subsubsec:LIB}, is used as the reflector locator (Figure~\ref{fig:all_models_overview}(a)).

Having knowledge from C-DYPSA about the first reflection TOAs $\hat{n}_{i,j}$, and assuming the vector containing the microphone position coordinates $\textbf{A}_i$ is known, multilateration generates spheres having radii equal to the TOAs $\hat{n}_{i,j}$ and centered at the respective sensor positions $\textbf{A}_i$. 
Minimizing a particular cost function which incorporates each reflection distance~\cite{BecStoLi2008}, the image source $\textbf{B}_{j}$ related to the $j$-th loudspeaker is obtained. 
However, with traditional multilateration, if microphones are too close to each other, $\textbf{B}_j$ cannot always be localized. Due to small errors during the TOA estimation, there are cases where the spheres do not intersect. Therefore, being unreliable with compact microphone arrays, the method was modified to randomly select three spheres, finding the point $\textbf{B}_{j,s}$, where $s$ indicates the selected three-microphone combination, and testing it for 100 combinations. When the algorithm fails, the combination is discarded. Thus, $S\leq 100$ potential image sources are found. The image position is taken as the mean over all the valid combinations: $\textbf{B}_{j}=\frac{1}{S}\sum_{s=1}^S \textbf{B}_{j,s}$.

\subsection{The Loudspeaker-Image Bisection (LIB) Algorithm}
\label{subsubsec:LIB}
The LIB algorithm was employed to localize the reflector by both \cite{TervoTossa2012} and \cite{DokParWalLuVet2013}, as shown in Figure \ref{fig:all_models_overview}(a).   
The plane $\textbf{p}_{j}$, defining the reflector, can be seen as the one bisecting the line $\textbf{l}_{j}$ from the $j$-th loudspeaker $\textbf{B}_{0,j}$ to the image $\textbf{B}_{j}$. Their midpoint $\textbf{M}_{j}$ lies on the plane. First, the unit vector normal to $\textbf{p}_{j}$ is defined:
\begin{equation}
\textbf{v}_{j}=\frac{\textbf{B}_{0,j}-\textbf{B}_{j}}{\|\textbf{B}_{0,j}-\textbf{B}_{j}\|}=[v_{1,j},v_{2,j},v_{3,j}]^T,
\label{eq:unit_vector_line}
\end{equation} 
where $\|\cdot \|$ stands for the Euclidean norm. Hence, this plane is defined in homogeneous coordinates as:
\begin{equation}
\begin{aligned}
\textbf{p}_{j}&=[\textbf{v}_{j}^T,~-\textbf{M}_{j}^T\textbf{v}_{j}]^T,
\end{aligned}
\label{eq:plane_from_normal_vector}
\end{equation}
where the midpoint is:
\begin{equation}
\textbf{M}_{j}\equiv \frac{\textbf{B}_{0,j}+\textbf{B}_{j}}{2}=[M_{1,j},M_{2,j},M_{3,j}]^T. 
\label{eq:mid-point}
\end{equation}

\section{Proposed Methods}
\label{sec:proposes_methods}

\subsection{Image-Source Reversion Methods}
\label{subsec:Image-Source Reversion Methods}
The proposed method ISDAR-LIB (Figure \ref{fig:all_models_overview}(a)) utilizes the same algorithm as~\cite{TervoTossa2012} and \cite{DokParWalLuVet2013} for the reflector estimation part (i.e. LIB), together with our image source locator ISDAR \cite{RemJackCol2015AES}. ISDAR exploits both TOAs and DOAs instead of TOAs only, as was previously done in \cite{TervoTossa2012} and \cite{DokParWalLuVet2013}. The two novel ISDAR-LIB variants (Figure \ref{fig:all_models_overview}(b)) are mean-ISDAR-LIB and median-ISDAR-LIB. Their main novelty is the integration of multiple loudspeakers. The ``multiple loudspeaker combination'' block, in Figure \ref{fig:all_models_overview}(b), represents mean and median, respectively. These two averages provide insight to the error types that most degrade localization performance: median is more robust to outliers rejecting all samples except the central one; whereas the mean is more robust to additive noise, reducing noise variance by $L$ for $L$ estimates.

\begin{algorithm}[t]
\caption{The ISDAR-LIB method}\label{alg:ISDAR-LIB}
\begin{algorithmic}[1]
\Statex \textbf{Input} TOAs $\overline{n}_{j}$ and DOAs $\Gamma_{j}=[\Theta_{j},\Phi_{j}]$ w.r.t. source $\textbf{B}_{0,j}$ and the microphone array center; source position
\Statex \textbf{Output} Plane $\textbf{p}_{j}$ (reflector)
\Statex \textit{/* ISDAR}
\State Calculate the radial distance $\overline{\rho}_{j}$ from $\overline{n}_{j}$
\State Localize the image source $\textbf{B}_{j}$ through Equation~(\ref{eq:source_position})
\Statex \textit{/* LIB}
\State Calculate the unit vector $\textbf{v}_{j}$ through Equation~(\ref{eq:unit_vector_line})
\State Calculate the midpoint $\textbf{M}_{j}$ of $\textbf{B}_{0,j}$ and $\textbf{B}_{j}$ (Equation~(\ref{eq:mid-point}))
\State Calculate the position of $\textbf{p}_{j}$ through Equation~(\ref{eq:plane_from_normal_vector})
\end{algorithmic}
\end{algorithm}

\begin{algorithm}[t]
\caption{The Mean-ISDAR-LIB method}\label{alg:Mean-ISDAR-LIB}
\begin{algorithmic}[1]
\Statex \textbf{Input} TOAs ${n}_{j}$ and DOAs $\Gamma_{j}=[\Theta_{j},\Phi_{j}]$ w.r.t. $L$ sources and the microphone array center; the source positions
\Statex \textbf{Output} Plane $\overline{\textbf{p}}_{\mathrm{M}}$ (reflector)
\For{$j\gets 1, L$}
\Statex ~~~~\textit{/* ISDAR-LIB}
\State Points between 1 and 4 in Algorithm \ref{alg:ISDAR-LIB}
\EndFor
\Statex \textit{/* Multiple loudspeaker based post-processing}
\State Calculate unit vector mean $\overline{\textbf{v}}$ from Equation~(\ref{eq:avg-LIB_components}) 
\State Calculate midpoint mean $\overline{\textbf{M}}$ from Equation~(\ref{eq:avg-LIB_components}) 
\State Calculate the plane $\overline{\textbf{p}}_{\mathrm{M}}$ utilizing $\overline{\textbf{v}}$ and $\overline{\textbf{M}}$
\end{algorithmic}
\end{algorithm}

\begin{algorithm}[t]
\caption{The Median-ISDAR-LIB method}\label{alg:Median-ISDAR-LIB}
\begin{algorithmic}[1]
\Statex \textbf{Input} TOAs ${n}_{j}$ and DOAs $\Gamma_{j}=[\Theta_{j},\Phi_{j}]$ w.r.t. $L$ sources and the microphone array center; the source positions
\Statex \textbf{Output} Plane $\overline{\textbf{p}}_{\mathrm{LS}}$ (reflector)
\For{$j\gets 1, L$}
\Statex ~~~~\textit{/* ISDAR-LIB}
\State Points between 1 and 4 Algorithm \ref{alg:ISDAR-LIB}
\EndFor
\Statex \textit{/* Multiple loudspeaker based post-processing}
\State Points 3 and 4 in Algorithm \ref{alg:Mean-ISDAR-LIB}
\State Calculate median midpoint as in Equation~(\ref{eq:median_midpoint_and_normal_vec})
\State Calculate median normal vector as in Equation~(\ref{eq:median_midpoint_and_normal_vec})
\State Calculate the plane $\widetilde{\textbf{p}}_{\mathrm{MED}}$ utilizing $\widetilde{\textbf{v}}$ and $\widetilde{\textbf{M}}$
\end{algorithmic}
\end{algorithm}

\subsubsection{Image Source Direction and Range (ISDAR) - LIB Method}
\label{subsubsec:ISDAR}
The selected baselines exploited information given by TOAs only. To improve the image source localization part, it was necessary to introduce an algorithm exploiting information from both TOAs (from C-DYPSA) and DOAs (from DSB). ISDAR is based on spherical coordinates, and we proposed a preliminary version in \cite{RemJackCol2015AES}. 
Given the radial distance $\overline{\rho}_{j}=\overline{n}_{j} c_0/F_s$, where $c_0$ is the sound speed and $F_s$ the sampling frequency, the azimuth $\Theta_{j}$ and the elevation $\Phi_{j}$, the image position $\textbf{B}_{j}=[x_{j},y_{j},z_{j}]^T$ can be written as: 
\begin{equation}
\begin{aligned}
&x_{j}=\overline{\rho}_{j}\cos(\Theta_{j})\cos(\Phi_{j}), \\
&y_{j}=\overline{\rho}_{j}\sin(\Theta_{j})\cos(\Phi_{j}), \\
&z_{j}=\overline{\rho}_{j}\sin(\Phi_{j}).
\end{aligned}
\label{eq:source_position}
\end{equation}

The reflector locator exploited by ISDAR-LIB is the same as the one utilized by the image-source reversion baselines \cite{TervoTossa2012}, \cite{DokParWalLuVet2013}, i.e. the LIB algorithm, already described in Section \ref{subsubsec:LIB}. Therefore, the plane estimated is $\textbf{p}_j$ from Equation~(\ref{eq:plane_from_normal_vector}). The pseudocode of ISDAR-LIB is reported in Algorithm~\ref{alg:ISDAR-LIB}.

\subsubsection{Mean-ISDAR-LIB} 
To improve the results provided by ISDAR-LIB, the information about the reflector position carried by multiple loudspeakers can be exploited. For this reason, mean-ISDAR-LIB is also proposed. It applies a multiple loudspeaker mean based post-processing algorithm to ISDAR-LIB. Considering $L$ loudspeakers, the $L$ midpoints $\textbf{M}_{j}$ and $L$ normal vectors $\textbf{v}_{j}$ can be obtained by ISDAR-LIB. The mean midpoint and the mean normal vector are calculated as:
\begin{equation}
\overline{\textbf{M}}=\frac{1}{L}\sum_{j=1}^L \textbf{M}_{j};~~~~~\overline{\textbf{v}}=\frac{1}{L}\sum_{j=1}^L \textbf{v}_{j}.
\label{eq:avg-LIB_components}
\end{equation}
Substituting $\overline{\textbf{M}}$ and $\overline{\textbf{v}}$ into Equation~(\ref{eq:plane_from_normal_vector}), $\overline{\textbf{p}}_{\mathrm{M}}$ is obtained. The pseudocode of mean-ISDAR-LIB is reported in Algorithm \ref{alg:Mean-ISDAR-LIB}.  

\subsubsection{Median-ISDAR-LIB} 
Another way to exploit multiple loudspeaker information from the $L$ estimated planes is to generate their median. Median-ISDAR-LIB calculates the median midpoint and median normal vector, from the ones obtained by applying ISDAR-LIB to $L$ loudspeakers, to generate $\widetilde{\textbf{p}}_{\mathrm{MED}}$. These are found as the midpoint $\widetilde{\textbf{M}}$ and normal vector $\widetilde{\textbf{v}}$ that are closest with respect to $\overline{\textbf{M}}$ and $\overline{\textbf{v}}$ (from Equation~(\ref{eq:avg-LIB_components})): 
\begin{equation}
\widetilde{\textbf{M}} = \argmin_{\textbf{M}_{j}}\|\overline{\textbf{M}}-\textbf{M}_{j}\|;~~~~~\widetilde{\textbf{v}} = \argmin_{\textbf{v}_{j}}\|\overline{\textbf{v}}-\textbf{v}_{j}\|.
\label{eq:median_midpoint_and_normal_vec}
\end{equation}
$\widetilde{\textbf{p}}_{\mathrm{MED}}$ is obtained by applying $\widetilde{\textbf{M}}$ and $\widetilde{\textbf{v}}$ in Equation~(\ref{eq:plane_from_normal_vector}). Pseudocode of median-ISDAR-LIB is given in Algorithm~\ref{alg:Median-ISDAR-LIB}.

\begin{figure}[t]
\centering
\includegraphics[trim={0pt 280pt 0pt 325pt},clip,width=0.5\columnwidth]{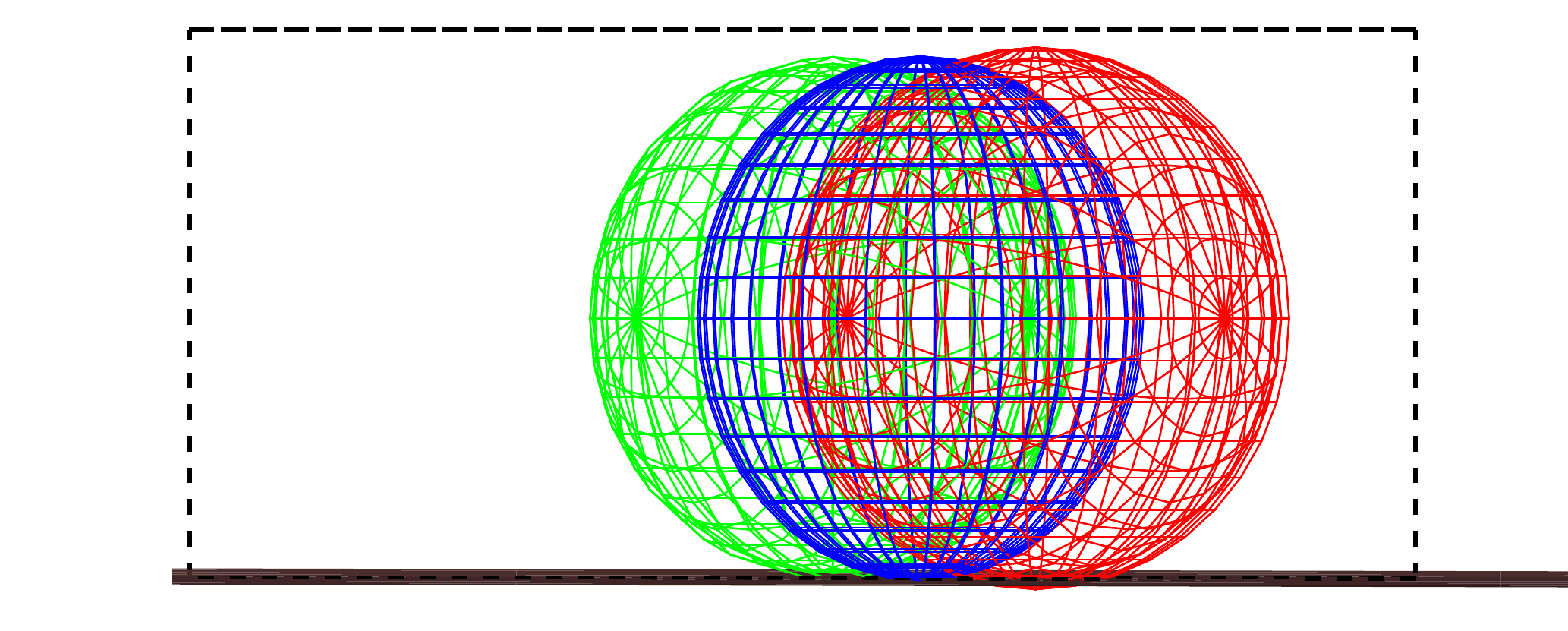}\hspace{-10pt}\includegraphics[trim={0pt 280pt 5pt 325pt},clip,width=0.5\columnwidth]{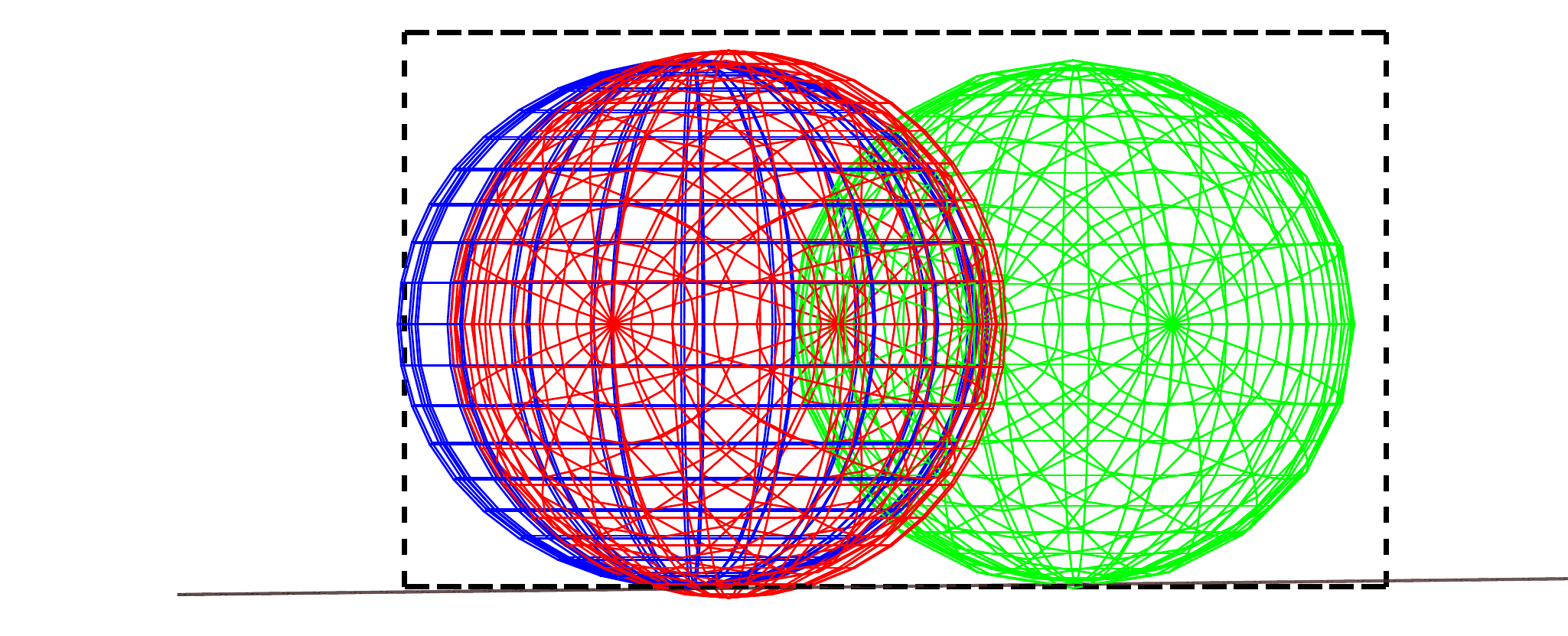}
\caption{Illustration of two elevation views showing the floor (solid brown line) estimated through ETSAC, the room groundtruth (dashed black lines), and ellipsoids constructed for three loudspeakers (red, green, blue).}
\label{fig:ETSAC_visualization}
\end{figure}

\subsection{Direct Localization Method}
\label{subsec:TDOA_reconstruction}
\subsubsection{Ellipsoid Tangent Sample Consensus (ETSAC)}
\label{subsubsec:ETSAC}
The proposed ETSAC (Figure \ref{fig:all_models_overview}(c)) is a direct localization method: it only has a reflector localization step. It uses the information extracted from multiple loudspeakers, and it differs from our recent work \cite{RemJackWangChamb2015} in the utilization of C-DYPSA as TOA estimator, instead of DYPSA. Here, RIRs recognized by C-DYPSA as outliers are not included by ETSAC in the analysis for the reflector localization. ETSAC first generates ellipsoids in homogeneous coordinates, having major axis equal to the respective reflection path, and foci on the loudspeaker-microphone combination. Then, RANSAC searches for the reflector~\cite{Fischler1981}. In other state-of-the-art methods based on ellipsoids (e.g. \cite{NasAntSarTub2011}), no specific TOA estimators are identified, and the reflector is informed through a cost function minimization. An example of the ETSAC output is shown in Figure \ref{fig:ETSAC_visualization}. Although in general ETSAC provides a unique solution, the particular case shown, with every microphone and loudspeaker at the same height, produces an up-down ambiguity. Prior knowledge may be used to constrain the solution (e.g., the floor is closer than the ceiling).

The ellipsoid has the property that the sum of the distances between a random point on its surface and its foci is constant. The TOA of the reflection yields the length of the reflection path. For this reason, ellipsoids having major axis equal to the reflection paths and foci on the respective microphone and source positions are constructed. By finding their common tangent plane, the reflector position is estimated. 
The parameters characterizing a general quadratic surface can be placed in a $4\times4$ matrix $\textbf{E}$, to define it in homogeneous coordinates:
\begin{equation}
\label{eq:3D C_matrix}
\textbf{E} = \left[\begin{matrix} a & d & f & g \\ d & b & e & h \\ f & e & c & i \\ g & h & i & j \end{matrix}\right],~~~\begin{vmatrix} a & d & f \\ d & b & e \\ f & e & c \end{vmatrix}>0.
\end{equation}
To represent a valid quadratic surface it has to satisfy: 
\begin{equation}
\label{eq:3D prop_matr}
det(\textbf{E})\neq0,~~~det(\textbf{E})/(a+b+c)<0.
\end{equation}
A unit sphere centered on the origin of the system is defined as the matrix $\textbf{E}_I$, obtained from Equation~(\ref{eq:3D C_matrix}) by choosing $a=b=c=1$, $j=-1$, and setting all the other coefficients equal to 0.
Transformations of translation, rotation and scaling are applied to model the ellipsoid with the required focus positions, axis directions and lengths~\cite{AkeHaiHoff}.
The sphere center is translated to the point $\Delta \textbf{X}_{i,j} = [\Delta x_{i,j},\Delta y_{i,j},\Delta z_{i,j}]^T$, through the translation matrix $\textbf{T}_{i,j}$.  
Assuming the source position $\textbf{B}_{0,j}=[x_{0,j},y_{0,j},z_{0,j}]^T$ is known, and the $i$-th microphone is located at $\textbf{A}_{i}=[A_{x,i},A_{y,i},A_{z,i}]^T$, we have $\Delta\textbf{X}_{i,j}= (\textbf{B}_{0,j}+\textbf{A}_i)/2. $
The major axis is defined as $Q^{maj}_{i,j}= \rho_{i,j}$, whereas the two minor axes are identical and coincide with $Q^{min}_{i,j}=\sqrt{\rho_{i,j}^2-\rho_{0,i,j}^2}$, where $\rho_{i,j}$ and $\rho_{0,i,j}$ are the path lengths relative to the reflection and direct sound respectively. The scaling matrix $\textbf{S}_{1,i,j}$ enlarges or shrinks the sphere utilizing $Q^{min}_{i,j}$ and $Q^{maj}_{i,j}$.
Finally, a 3D rotation matrix $\textbf{R}_{i,j}$ is generated utilizing the angles of rotation $\alpha_{i,j} = \mathrm{arctan}\left(\frac{z_{0,j}-A_{z,i}}{y_{0,j}-A_{y,i}}\right)$, $\beta_{i,j}= \mathrm{arctan}\left(\frac{x_{0,j}-A_{x,i}}{z_{0,j}-A_{z,i}}\right)$, and $\gamma_{i,j}= \mathrm{arctan}\left(\frac{y_{0,j}-A_{y,i}}{x_{0,j}-A_{x,i}}\right)$.
Therefore, the matrix defining the $i$-th microphone and $j$-th loudspeaker ellipsoid is: 
\begin{equation}
\textbf{E}_{i,j} = \textbf{T}_{i,j}^{-T}\textbf{R}_{i,j}^{-T}\textbf{S}_{i,j}^{-T}\textbf{E}_I\textbf{S}_{i,j}^{-1}\textbf{R}_{i,j}^{-1}\textbf{T}_{i,j}^{-1}.
\label{eq:ellipsoids}
\end{equation}

\begin{algorithm}[t]
\caption{The ETSAC method}\label{alg:ETSAC}
\begin{algorithmic}[1]
\Statex \textbf{Input} TOAs ${n}_{0,i,j}$ and ${n}_{i,j}$ (direct sound and reflection paths, respectively) w.r.t. the $L$ sources and the $M$ microphones; every source $\textbf{B}_{0,j}$ and microphone $\textbf{A}_{i}$ positions 
\Statex \textbf{Output} Plane $\textbf{p}_{\mathrm{ETSAC}}$ (reflector)
\For{$i\gets 1, M$}
\For{$j\gets 1, L$}
\State The unit sphere $\textbf{E}_I$
\State The distances ${\rho}_{0,i,j}$ and ${\rho}_{i,j}$ from ${n}_{0,i,j}$ and ${n}_{i,j}$
\State The parameters $\Delta\textbf{X}_{i,j}$,$Q^{maj}_{i,j}$, $Q^{min}_{i,j}$, $\alpha_{i,j}$, $\beta_{i,j}$, $\gamma_{i,j}$ 
\State The matrices $\textbf{T}_{i,j}$, $\textbf{R}_{i,j}$ and $\textbf{S}_{i,j}$
\State The ellipsoid through Equation~(\ref{eq:ellipsoids})
\EndFor
\EndFor
\For{$p\gets 1, P$}
\State The random unit vector $\textbf{v}_{p}$
\State The plane $\textbf{p}_{p}$ through $\textbf{v}_{p}$ and Equation~(\ref{eq:v1_v2_v3})
\For{$n\gets 1, N=L\cdot M$}
\State The tangency coefficient $t_{n,p}$
\If{$t_{n,p}<\tau_t$}
\State The $n$-th ellipsoid is considered tangent
\EndIf
\EndFor
\EndFor
\State The $p$-th plane with the most tangent ellipsoids, and lowest tangency coefficients $t_p$, is $\textbf{p}_{\mathrm{ETSAC}}$
\end{algorithmic}
\end{algorithm}

Once all the $N=L\cdot M$ ellipsoids are defined, where $L$ is the number of loudspeakers and $M$ the number of microphones in the array, the next step is to find their common tangent plane. The approach is to randomly select a certain number of points $P$ on the ellipsoid with coefficients $i=1$ and $j=1$, and verify, by setting a threshold, which one generates the plane closest to the required one. Having randomly generated a normal vector $\textbf{v}_{p}=[v_{1,p},v_{2,p},v_{3,p}]^T$, the $p$-th plane tried during the algorithm can be defined in homogeneous coordinates, following Equation~(\ref{eq:plane_from_normal_vector}), as $\textbf{p}_p=[v_{1,p},v_{2,p},v_{3,p},p_{4,p}]^T$. The coefficient $p_{4,p}$ can be calculated by considering the general property of tangency between a plane and an ellipsoid, $\textbf{p}^T_{p}\textbf{E}^*_{1,1}\textbf{p}_{p}=0$, where $\textbf{E}_{1,1}^*$ is the adjoint matrix of $\textbf{E}_{1,1}$. A system of four random ellipsoid equations is then constructed to obtain $p_{4,p}=(-w_2+\sqrt{w_2^2-4w_1w_3})/2w_1$, where:
\begin{equation}
\begin{aligned}
w_1 &= j^*, \\
w_2 &= 2(g^*v_{1,p} + h^*v_{2,p} + i^*v_{3,p}), \\
w_3 &= a^*v_{1,p}^2 + b^*v_{2,p}^2 + c^*v_{3,p}^2 + \\
    &+ 2(d^*v_{1,p}v_{2,p} + e^*v_{2,p}v_{3,p} + f^*v_{1,p}v_{3,p}),
\end{aligned}
\label{eq:v1_v2_v3}
\end{equation}
where $a^*, b^*, c^*, d^*, e^*, f^*, g^*, h^*, i^*$ and $j^*$ are the elements of the matrix $\textbf{E}^*_{1,1}$, organized in the same order as for the general matrix in Equation~(\ref{eq:3D C_matrix}). To verify if the plane is tangent to the $N$ ellipsoids, the tangency coefficient is calculated for each of them as $t_{n,p}=|\textbf{p}^T_{p}\textbf{E}^*_{i,j}\textbf{p}_{p}|$, where $|\cdot |$ indicates the absolute value.
Since the $p$-th plane is perfectly tangent to the $n$-th ellipsoid if $t_{n,p}=0$, a threshold $\tau_t$ is heuristically set depending on the dataset used and, when $t_{n,p}>\tau_t$, the $n$-th ellipsoid is considered non-tangent. The plane with the fewest non-tangent ellipsoids is selected as the estimated $\textbf{p}_{\mathrm{ETSAC}}$. In the scenario where more than one plane has the fewest non-tangent ellipsoids, the plane with the lowest sum of tangency coefficients $t_p=\sum_{n=1}^N t_{n,p}$ is selected as $\textbf{p}_{\mathrm{ETSAC}}$. The ETSAC pseudocode is shown in Algorithm~\ref{alg:ETSAC}.

\begin{table}[!t]

\caption{Parameter values used for the experiments.}
\label{tab:reproducibility_params}
\centering

\begin{tabular}{|c|c|c|}
\hline
\textbf{Parameter name} & \textbf{Symbol} & \textbf{Value} \\
\hline
Slope function threshold (C-DYPSA) & $\tau_S$ & 0.2 \\
\hline
Amplitude threshold (C-DYPSA) & $\tau_A$ & 25dB \\
\hline
Group-delay window length (C-DYPSA) & $T_{GD}$ & $3.5\cdot 10^{-3}\,s$  \\
\hline
Segmentation window length (DSB) & $T$ & $2.7\cdot 10^{-3}$\,$\mathrm{s}$  \\ 
\hline
Space samples (ML)  & $x$ & $10^4$  \\
\hline
RANSAC samples (ETSAC) & $P$ & $10^4$  \\
\hline
RANSAC threshold (ETSAC) & $\tau_t$ & $1.4\cdot 10^{-3}$  \\
\hline

\end{tabular}
\end{table}

\section{Experimental evaluation and discussion}
\label{sec:exp_eval}
In this section, we evaluate the proposed methods and compare them with the baseline methods~\cite{TervoTossa2012, DokParWalLuVet2013}, fulfilling the last contribution of this article, defined in Section \ref{sec:introduction}. We first describe how the data used in our experiments were generated or recorded, and then discuss the performance metrics, before presenting the comparative studies in terms of reflector localisation accuracy and computational cost. The C-DYPSA performance is also compared to DYPSA~\cite{NayKouGudBroo}. 

As no other RIR dataset was publicly available for microphone arrays that can be defined as compact (see Section \ref{subsec:assumptions}), we decided to simulate and record the data for our experiments. A 48-channel bi-circular array with a typical microphone spacing of 21\,mm (spatial aliasing half wavelength at 8\,kHz) and an aperture of 212\,mm (wavelength at 400\,Hz) was deployed in four rooms with different sizes and reverberation times (RT60s). This data, recorded with a sampling rate of $F_s=48$\,kHz, is available online at \cite{MCRIRdataset2015}\footnote{http://cvssp.org/data/s3a/}. The experiments were run on MATLAB R2014b on Intel(R) Core(TM)i7-2600 CPU @ 3.40GHz, 16GB RAM PC. To aid the reproducibility, the values of the parameters used, heuristically obtained for the employed datasets, are reported in Table \ref{tab:reproducibility_params}.

\begin{table}[!t]

\caption{Room dimensions (m), and volumes ($\mathrm{m}^3$) in brackets, for the 10 simulated rooms. When the absorption coefficient $\overline{\eta}=0.5$, 2 medium-sized rooms were simulated.}
\label{tab:simulated_datasets}
\centering

\resizebox{\columnwidth}{!}{%
\begin{tabular}{|c|c|c|c|}
\cline{2-4}
\multicolumn{1}{c|}{} & $\overline{\boldsymbol\eta}\boldsymbol{=}\mathbf{0.2}$ & $\overline{\boldsymbol\eta}\boldsymbol{=}\mathbf{0.5}$ & $\overline{\boldsymbol\eta}\boldsymbol{=}\mathbf{0.8}$ \\
\hline
\textbf{Small} & 6.0,\,4.3\,,2.3\,(59) & 2.4,\,4.0,\,2.4\,(23) & 4.1,\,5.0,\,2.1\,(43)  \\
\hline
\textbf{Medium} & 7.4,\,5.7,\,2.5\,(105) & \specialcell{7.4,\,5.7,\,2.5\,(105)\\7.8,\,6.1,\,4.0\,(189)} & 7.4,\,5.7,\,2.5\,(105) \\
\hline
\textbf{Large} & 19.7,\,24.3,\,6.0\,(2872) & 14.6,\,17.1,\,6.5\,(1623) & 6.6,\,8.8,\,4.0\,(232)  \\
\hline

\end{tabular}
}
\end{table}

\subsection{Datasets}
\label{subsec:datasets}

\subsubsection{Simulated Datasets}
\label{subsubsec:simulated_datasets}
Ten rooms were simulated, with varying dimensions and absorption coefficients covering a typical range. They are classified by size and average absorption coefficient $\overline{\eta}$ in Table \ref{tab:simulated_datasets}. Inside each room, ten different loudspeaker and microphone array configurations were randomly chosen, leading to a total of 100 different setups. The image source model was employed to generate RIRs, through a Matlab toolbox~\cite{Habets}. The maximum order of the reflections was set to 5, and the high-pass filter, employed to eliminate the artificial energy at the low frequencies, was enabled. 
The loudspeakers were randomly positioned on a circle around the center of the microphone array, following a uniform distribution over azimuthal angles, not allowing interspaces between the loudspeakers of less than 5 degrees. Two radii of the circle were chosen: 1.00\,m for the small sized rooms, and 1.68\,m for the medium and large rooms. Their height was the same as for the microphone array, i.e. 0.90\,m. The simulated microphone array was composed by 48 evenly spaced microphones placed in two concentric circles, with the inner circle of radius 0.083\,m, and outer circle of 0.104\,m radius, similar to the prototype designed for our experimental apparatus. Its circular configuration was chosen since it has been proved to be effective for analyzing acoustic 3D information \cite{deVHorGro2007}. The center of the microphone array on the horizontal plane was randomly chosen. However, a limit was set to maintain the loudspeakers at a minimum distance from the reflectors. For the small rooms this distance was set to be 0.22\,m, whereas for the medium and large rooms 0.36\,m. 
Two noise regimes were imposed on the simulated RIRs, to examine the effects of microphone misplacement~\cite{RemJackCol2015ICSV} and additive measurement noise, respectively. For the first regime, spatial vectors were generated and applied to modify the original microphone positions. They had random directions and amplitudes: the maximum amplitude was 7\,mm, i.e., 1 sample at sampling frequency $F_s=48$\,kHz. This regime also models systematic bias in propagation uncertainty.  Independent white Gaussian noise was added to each RIR, providing a direct-to-noise ratio (DNR) of 70\,dB. 

For the second regime, extra datasets were generated with the same ten rooms, randomly choosing ten loudspeaker and microphone array configurations for each, as described above. In this case, we wanted to observe the performance of the methods in the scenario where the acoustic channel is estimated \cite{NayGau2010}. The maximum amplitude for the microphone displacement was 1\,mm, and the DNR was set to be either 30\,dB, 40\,dB, or 50\,dB. Therefore, having 100 room setups for every DNR, there were a total of 300 additional simulated datasets.

\begin{figure}[t]
\centering
\includegraphics[trim={0pt 0pt 0pt 0pt},clip,width=1\columnwidth]{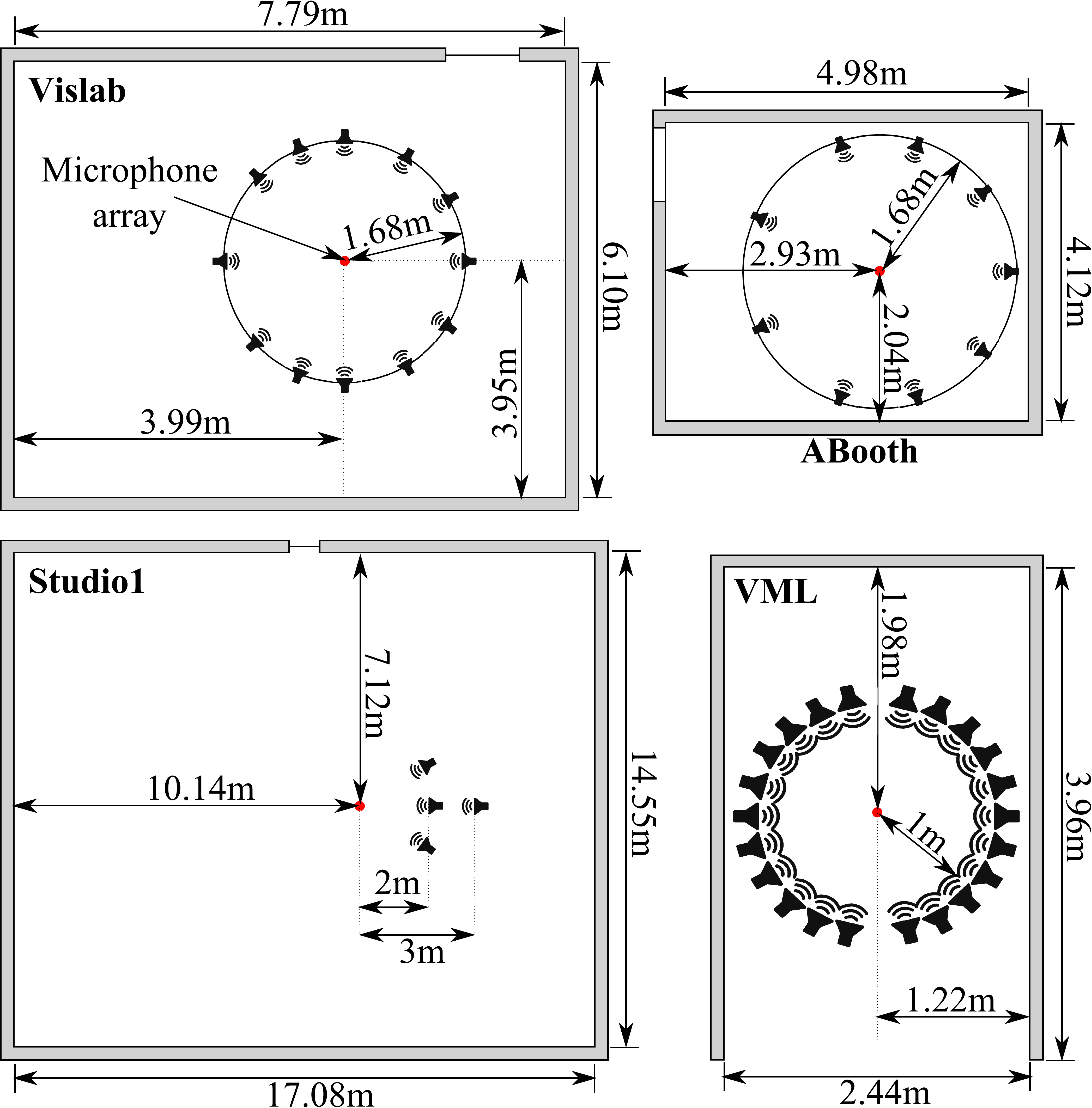}
\caption{Floor plan view of the four measured rooms (different scales). The loudspeaker (loudspeaker symbol) and microphone array (red circle) positions are also illustrated. The ceilings are at 2.10\,m in AudioBooth, 3.98\,m in Vislab, 2.42\,m in VML, and 6.50\,m in Studio1.}
\label{fig:Plan_rooms}
\end{figure}

\subsubsection{Recorded Datasets}
\begin{figure}[t]
\centering
\includegraphics[width=\columnwidth]{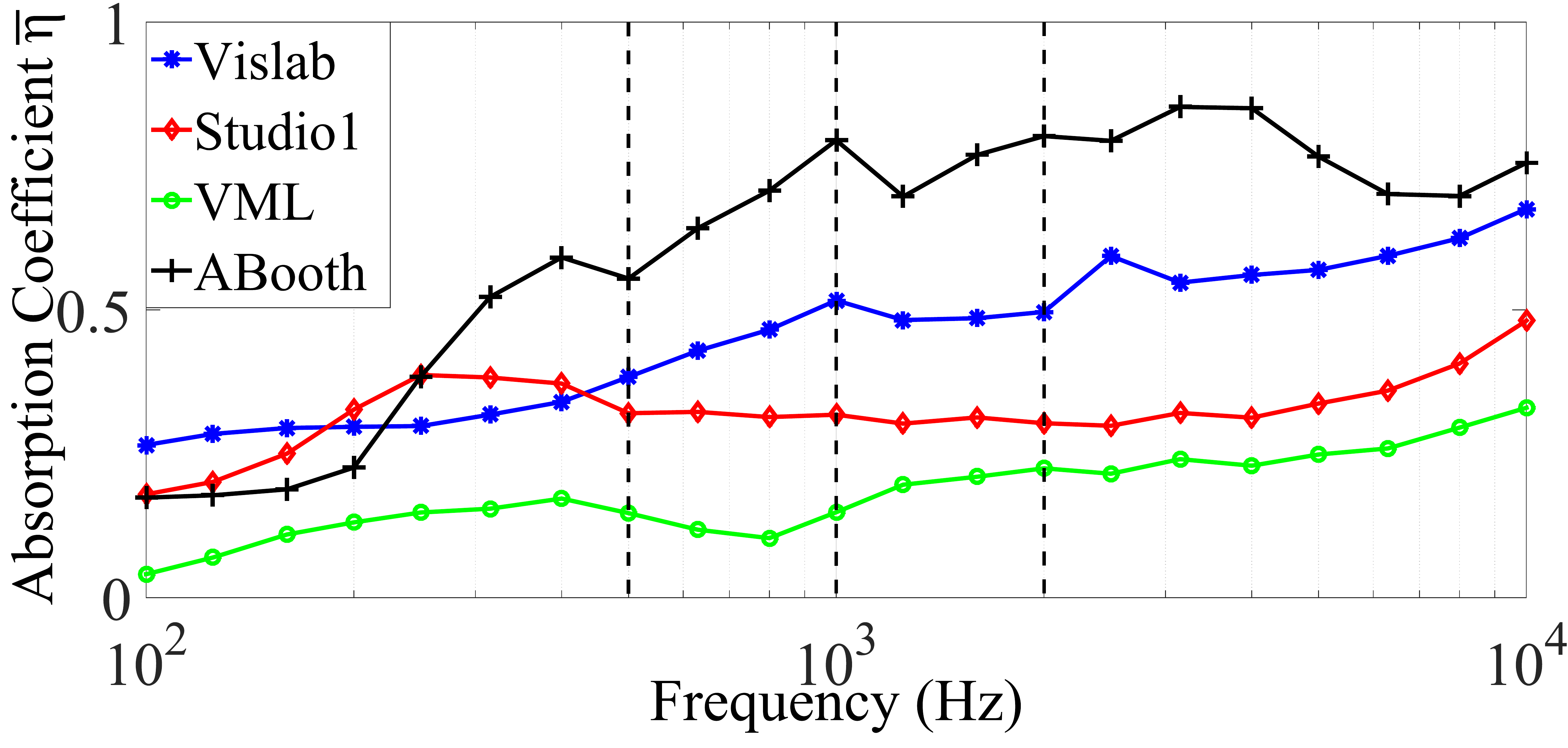}
\caption{The average absorption coefficient $\overline{\eta}$ of the recorded rooms, evaluated in $\frac{1}{3}$-octave bands. 500\,Hz, 1\,kHz and 2\,kHz are highlighted with dotted vertical lines.}
\label{fig:absorption_coefficients}
\end{figure}

The rooms recorded are named as ``Vislab'', ``Studio1'', ``AudioBooth'' (ABooth) and ``VML'', and plan view of each of them is shown in Figure \ref{fig:Plan_rooms}. Table \ref{tab:datasets} reports their general acoustic characteristics, including the average absorption coefficient $\overline{\eta}$ in third octave bands. Their $\overline{\eta}$s are also shown over the range between 100\,Hz and 10\,kHz, in Figure \ref{fig:absorption_coefficients}. This is calculated by the inverse of Sabine's equation $\overline{\eta} \approx (0.161\cdot V)/(S\cdot \mathrm{RT}60)$, where $V$ is the room volume, and $S$ is the total reflective surface area~\cite{Kuttruff4}. These rooms were chosen since they cover ranges between small and large $V$, and between small and large $\overline{\eta}$ \cite{LinKosWei2012}. ``Vislab'' can be considered as characterized by a medium $V$ (suitable for 20 people) and large $\overline{\eta}$, ``Studio1'' by a large $V$ (for 200 people) and medium $\overline{\eta}$, whereas ``VML'' has both small $V$ (for 2 people) and $\overline{\eta}$. In addition, ``AudioBooth'' is characterized by a small $V$ (for 2 people), and a peculiar $\overline{\eta}$, which is very large for high frequencies and medium for low frequencies. Every dataset was recorded using the swept sine RIR method, and the sound speed assumed to be $c_0=343.1\,\mathrm{m}\cdot \mathrm{s}^{-1}$.
To analyze the methods varying only parameters like size and RT60, similar $M$ and $L$ must be chosen, therefore, subsets of these datasets were selected. In addition, to be uniform across the datasets, for every room except ``VML'', loudspeakers were selected in the horizontal plane only, with the same height as the microphones. In every room the same 48 channel bi-circular compact uniform array of Countryman B3 omni microphones was used, similar to the design of the simulated array in Section~\ref{subsubsec:simulated_datasets}. However, there is a small discrepancy in the size of the array used for generating the simulated data and real recordings. We simulated the array considering the original design, although, due to manufacturer tolerances, the real one has a radius 2\,mm wider. Genelec 8020B loudspeakers were used. Although a description of the datasets is reported below, for further details, the reader can refer to \cite{RemJacColFra2015, LiuWanJacCox2015, MCRIRdataset2015}.     

\begin{table}[!t]

\caption{RIR dataset room properties: reverberation time RT60, Dimensions, volume $V$, average absorption coefficient $\overline{\eta}$, and number of loudspeakers used $L$.}
\label{tab:datasets}
\centering

\resizebox{\columnwidth}{!}{%
\begin{tabular}{|c|c|c|c|c|}
\hline
\textbf{Dataset} & \specialcell{\textbf{RT60 (ms)}\\\textbf{0.5--1--2\,kHz}} & \textbf{Dim. (m), (\textit{V}}) \textbf{(}$\bf{\mathrm{\textbf{m}}^3}$\textbf{)} & \specialcell{$\overline{\boldsymbol\eta}$\\\textbf{0.5--1--2\,kHz}} & \textbf{\textit{L}} \\
\hline
ABooth & 158--110--109 & 4.1,\,5.0,\,2.1\,(43) & 0.55--0.79--0.80 & 9 \\
\hline
Vislab & 385--286--306 & 7.8,\,6.1,\,4.0\,(189) & 0.38--0.51--0.50 & 12\\
\hline
VML & 505--499--330 & 2.4,\,4.0,\,2.4\,(23) & 0.15--0.15--0.22 & 22 \\
\hline
Studio1 & 894--901--945 & 14.6,\,17.1,\,6.5\,(1623) & 0.32--0.32--0.30 & 4 \\ 
\hline

\end{tabular}
}
\end{table}

The ``AudioBooth'' is an acoustically treated room at the University of Surrey \cite{RemJacColFra2015}. Nine loudspeakers were selected for this paper, lying around the equator of a truncated geodesic sphere, at 1.68\,m radius, at $0$, $\pm 30$, $\pm 70$, $\pm 110$ and $\pm 155$ degrees in azimuth relative to the center channel. The microphone array was positioned at the center of the sphere at a height of 1.02\,m. 
``Vislab'' is another acoustically treated room at the University of Surrey, where the ``Surrey Sound Sphere'', having radius of 1.68\,m, has been assembled \cite{RemJackWangChamb2015}. Twelve loudspeakers clamped on the sphere equator, at a height of 1.62\,m, with azimuth $0$, $\pm 30$, $\pm 60$, $\pm 90$, $\pm 110$, $\pm 135$ and $180$ degrees, were selected for this paper. The microphone array was placed at the center of the sphere. 
``VML'' is a mock room built within a lab at the University of Surrey, with one wall and ceiling missing like a film set \cite{LiuWanJacCox2015}. The microphone array was hanging, at the height of 2.20\,m, at the center of the room. 24 loudspeakers were laid equispaced around the array with 1\,m radius, and facing the center. The two loudspeakers equidistant from the two walls were discarded, introducing ambiguities with C-DYPSA, the common pre-processor of every tested method.     
``Studio1'' is a large recording studio at the University of Surrey \cite{RemJacColFra2015}. Four loudspeaker positions were used, at a height of 1.50\,m (the same used for the microphone array). Three of them were placed at a distance of 2\,m from the microphone array and azimuth of $0$ and $\pm 45$ degrees, whereas the fourth one was at 0 degrees azimuth and 3\,m distant.

\subsection{Evaluation Metrics}
\label{subsect:eval_metrics}
Large errors, generated during the process of finding the image source or the reflector position, can highly influence average results. This may not allow discrimination of smaller error behaviors. Therefore, a distinction was made between gross and fine errors, defining a threshold at 500\,mm as in~\cite{OmoSvaBruCri2006}. 

\subsubsection{TOA Estimation}
\label{subsubsec:TOA_Estimation}
For consistent evaluation in spatial terms, the TOA was evaluated as the corresponding propagation distance $\rho_{i,j}=n_{i,j}c_0/F_s$, where $n_{i,j}$ is the TOA in samples of the reflection path between the $i$-th microphone and the $j$-th loudspeaker, $c_0=343.1\,\mathrm{m\cdot s^{-1}}$ is the sound speed, and $F_s$ the sampling frequency. The error $\epsilon^{\mathrm{TOA}}_{i,j}$, is thus calculated as the distance (in mm) between $\rho_{i,j}$ and its groundtruth. With $L$ loudspeakers and $M$ microphones, for a total of $N$ combinations, the overall error is then obtained as the root mean square error (RMSE):
\begin{equation}
\label{eq:error_TOAs}
\mathrm{RMSE_{TOA}} = \sqrt{\frac{1}{N}\sum_{i=1}^M\sum_{j=1}^{L}(\epsilon^{\mathrm{TOA}}_{i,j})^2}.
\end{equation}

\subsubsection{Image Source Localization}
The image source localization errors $\epsilon_j$ were evaluated as the Euclidean distance between each single estimated image $\textbf{B}_{j}$ and its own groundtruth $\textbf{B}_{Gj}$, averaged over all the $X_I\leq L$ loudspeakers giving fine errors:
\begin{equation}
\label{eq:error_images}
\mu_\epsilon = \frac{1}{X_I}\sum_{j=1}^{X_I} \epsilon_j,~~~~~~~~~~\mathrm{for}~\epsilon_j<500\,\mathrm{mm}.
\end{equation}

\subsubsection{Reflector Localization}
To obtain the error in reflector positioning, $K=5$ equispaced points were selected between each of the $Y=C_1^{M(L-1)}$ source-sensor combinations. The projections of these $Z=K\cdot Y$ points on the estimated plane $\textbf{p}$ and the relative groundtruth $\textbf{p}_G$ were then calculated. The Euclidean distance between each pair of points is obtained to give the errors $e_z$. To provide a reliable measure, the RMSE was calculated over all $Z$ points as $\mathrm{RMSE}_l = \sqrt{\frac{1}{Z}\sum_{z=1}^Z e_z^2}$, indicating the error of each estimated plane. Then, to be coherent with the image source evaluation and provide a summary value for each dataset, the average was calculated over every plane. To do this, for all the reflector localization methods exploiting multiple loudspeakers (i.e. mean-ISDAR-LIB, median-ISDAR-LIB and ETSAC), the leave-one-loudspeaker-out method was applied. It consists of selecting $L-1$ loudspeakers, where $L$ are the loudspeakers in the dataset. All the combinations $U=C_{L-1}^L=L$ were tested, and the average over the $X_R\leq U$ ones with fine errors provided:
\begin{equation}
\label{eq:AVG_RMSE_reflectors}
\mu_{\mathrm{RMSE}} = \frac{1}{X_R}\sum_{l=1}^{X_R} \mathrm{RMSE}_l,~~~~\mathrm{for}~\mathrm{RMSE}_l<500\,\mathrm{mm}.
\end{equation}

\subsubsection{Confidence Interval and Gross Error}
\label{subsubsec:CI_G}
The gross error rates were evaluated as $G_\epsilon =(1-X_I/L)\cdot 100$ and $G_\mathrm{RMSE} =(1-X_R/U)\cdot 100$, together with their average over the different datasets and related confidence interval. In contrast to the outlier thresholds defined in Equations (\ref{eq:error_images}) and (\ref{eq:AVG_RMSE_reflectors}) for image source and reflector evaluation, the threshold separating gross and fine TOA estimation errors was set to match the maximum microphone distance from the array center, i.e., the array radius of 104\,mm. These gross TOA errors were named $G_{\mathrm{TOA}}$.

To provide a better statistical evaluation of the results, the confidence interval of the average across the datasets was calculated as~\cite{Rao2009}:
\begin{equation}
\label{eq:confidence_interval}
\mathrm{CI}_\epsilon = \zeta\frac{1}{D}\sqrt{\sum_{d=1}^{D} (\mu_{\epsilon,d}-\frac{1}{D}\sum_{d=1}^{D}\mu_{\epsilon,d})^2},
\end{equation}
where $\zeta=1.96$ is the critical value for a confidence interval of $95\%$, $D$ is the number of datasets available, and $d$ is the dataset index. 
The confidence interval for the RMSEs ($\mathrm{CI}_\mathrm{RMSE}$) was calculated, by substituting into Equation~(\ref{eq:confidence_interval}), $\mu_{\epsilon}$ with $\mu_{\mathrm{RMSE}}$; similarly for the TOA estimation confidence interval $\mathrm{CI}_\mathrm{TOA}$, by substituting in Equation~(\ref{eq:confidence_interval}) $\mu_{\epsilon}$ with $\mathrm{RMSE_{TOA}}$.

\begin{table}[!t]

\caption{The top part shows the gross error $G_{\mathrm{TOA}}$ for the first reflection. The bottom part shows $\mathrm{RMSE_{TOA}}$ and $\mathrm{CI_{TOA}}$, of the reflection path length calculated using DYPSA and C-DYPSA, for the four recorded datasets, expressed in mm. }
\label{tab:C-DYPSA_eval}
\centering

\resizebox{\columnwidth}{!}{%
\begin{tabular}{|c|c|c|c|c|c|}
\cline{2-6}
\multicolumn{1}{c|}{$\bm{G_{\mathrm{TOA}} (\%)}$} & \textbf{ABooth} & \textbf{Vislab} & \textbf{VML} & \textbf{Studio1} & \textbf{AVG} \\
\hline
\textbf{DYPSA \cite{NayKouGudBroo}} & $2.3$ & $15.5$ & $27.1$ & $0.5$ & $11.4\pm 10.6$ \\
\hline
\textbf{C-DYPSA} & $\bm{0.7}$ & $\bm{11.5}$ & $\bm{21.1}$ & $\bm{0.0}$ & $\bm{8.3\pm 8.5}$ \\
\hline
\hline
\multicolumn{1}{c|}{$\bm{\mathrm{RMSE_{TOA}}}$ \textbf{(mm)}} & \textbf{ABooth} & \textbf{Vislab} & \textbf{VML} & \textbf{Studio1} & \textbf{AVG} \\
\hline
\textbf{DYPSA \cite{NayKouGudBroo}} & $54$ & $110$ & $194$ & $100$ & $115\pm 50$ \\
\hline
\textbf{C-DYPSA} & $\bm{48}$ & $\bm{95}$ & $\bm{192}$ & $\bm{99}$ & $\bm{109\pm 51}$ \\
\hline
\end{tabular}
}
\end{table}

\subsection{C-DYPSA Evaluation}
\label{subsec:C-DYPSA_eval}
The novel TOA estimator C-DYPSA (Section \ref{subsubsec:C-DYPSA}) was evaluated and compared against its previous version, the DYPSA algorithm \cite{NayKouGudBroo}, on the four recorded datasets. Experiments were performed calculating $\mathrm{RMSE_{TOA}}$ (Equation~(\ref{eq:error_TOAs})), and $\mathrm{CI}_\mathrm{TOA}$ (Equation~(\ref{eq:confidence_interval})). Results are reported in the bottom of Table \ref{tab:C-DYPSA_eval}. C-DYPSA performed better in every dataset, since outliers produced by DYPSA for single RIRs are discarded in C-DYPSA, generating a final estimate more robust and accurate. The top part of Table \ref{tab:C-DYPSA_eval} shows the gross errors $G_{\mathrm{TOA}}$ decreasing for every dataset, applying C-DYPSA.

This article concerns reflector localization methods with the first reflection; yet for other early reflections, the clustering of responses across microphones by C-DYPSA can exploit the array's compactness to reduce errors in the association of epochs to a reflection. For later higher-order reflections, C-DYPSA fails predictably as the level of the reflection energy falls towards the noise floor. A further study could investigate how the number of detectable early reflections varies with the quality of the recordings, and the room properties. Here, C-DYPSA is used to clean up the input to the reflector localization methods. In addition, we have performed experiments to compare DYPSA with the state-of-the-art algorithm in \cite{Kuster2008}. We observed that the method in \cite{Kuster2008} had considerably lower performance in estimating TOAs, and therefore, the results for this method are not included in this paper.   

\begin{table}[!t]

\caption{Reflector localization gross errors $\bm{G_{\mathrm{RMSE}}}$, averaged RMSE $\bm{\mu_{\mathrm{RMSE}}}$, and confidence interval $\bm{\mathrm{CI}_\mathrm{RMSE}}$, for the simulated dataset grouped by room size (Small, Medium, Large) and absorption coefficient $\bm{\overline{\eta}}$, with overall values. The methods are: (A) ISDAR-LIB, (B) median-ISDAR-LIB, (C) mean-ISDAR-LIB, and (D) ETSAC.}
\label{tab:simulated_reflector_localization}
\centering

\begin{tabular}{|c|c|c|c||c|c|c||c|}
\cline{2-7}
\multicolumn{1}{c|}{\textbf{$\bm{G_{\mathrm{RMSE}}}$}}& \multicolumn{3}{c||}{\textbf{Size}} & \multicolumn{3}{c||}{$\bm{\overline{\eta}}$} & \multicolumn{1}{c}{} \\ 
\cline{2-8}
\multicolumn{1}{c|}{$\bm{(\%)}$}& \textbf{S} & \textbf{M} & \textbf{L} & \textbf{$\bm{0.2}$} & \textbf{$\bm{0.5}$} & \textbf{$\bm{0.8}$} & \textbf{Overall} \\
\hline
\textbf{A} & 29.7 & 0.5 & \textbf{0.0} & 3.6 & 19.1 & 6.6 & $9.9\pm 8.7$\\
\hline
\textbf{B} & 14.0 & \textbf{0.0} & \textbf{0.0} & 2.0 & 11.4 & 1.0 & $4.7\pm 4.6$ \\
\hline
\textbf{C} & 9.7 & \textbf{0.0} & \textbf{0.0} & 1.0 & 8.2 & 1.0 & $3.3\pm 3.2$\\
\hline
\textbf{D} & \textbf{8.1} & \textbf{0.0} & \textbf{0.0} & \textbf{0.1} & \textbf{5.9} & \textbf{0.4} & $\bm{2.4\pm 2.6}$ \\
\hline
\hline
\multicolumn{1}{c|}{$\bm{\mu_{\mathrm{RMSE}}}$}& \multicolumn{3}{c||}{\textbf{Size}} & \multicolumn{3}{c||}{$\bm{\overline{\eta}}$} & \multicolumn{1}{c}{} \\ 
\cline{2-8}
\multicolumn{1}{c|}{\textbf{(mm)}}& \textbf{S} & \textbf{M} & \textbf{L} & \textbf{$\bm{0.2}$} & \textbf{$\bm{0.5}$} & \textbf{$\bm{0.8}$} & \textbf{Overall} \\
\hline
\textbf{A} & $61$ & $13$ & $13$ & $13$ & $31$ & $44$ & $29\pm 15$ \\
\hline
\textbf{B} & $27$ & $34$ & $34$ & $30$ & $29$ & $34$ & $31\pm 2$ \\
\hline
\textbf{C} & $207$ & $34$ & $24$ & $60$ & $151$ & $109$ & $98\pm 52$ \\
\hline
\textbf{D} & $145$ & $13$ & $14$ & $13$ & $107$ & $71$ & $61\pm 41$ \\
\hline

\end{tabular}
\end{table}

\subsection{Simulated Experiments}
Experiments were performed considering the simulated datasets introduced in Section \ref{subsubsec:simulated_datasets}. The aim of these simulations was to evaluate the proposed reflector localization methods, over a wide variety of controlled scenarios, highlighting potential strengths and weaknesses. The metrics utilized were $\mu_{\mathrm{RMSE}}$ (Equation~(\ref{eq:AVG_RMSE_reflectors})) and  $\mathrm{CI}_\mathrm{RMSE}$ (Equation~(\ref{eq:confidence_interval})) to evaluate the fine errors, and $G_\mathrm{RMSE}$ (Section \ref{subsubsec:CI_G}) for the gross errors. Two different sets of simulations were performed. First, the 100 datasets produced by varying size and $\overline{\eta}$, with direct sound 70\,dB louder than the additive noise, and microphone perturbation of 7\,mm maximum, were evaluated, with results reported in Table \ref{tab:simulated_reflector_localization}. Then, the 300 datasets obtained by varying the DNR were considered, and the results are shown in Table \ref{tab:simulated_SNR}. 

Starting from the first set of simulations (Table \ref{tab:simulated_reflector_localization}, top), the direct localization ETSAC gives the best performance, with the lowest $G_\mathrm{RMSE}$ over the 100 datasets. The multiple-loudspeaker methods (i.e. median-ISDAR-LIB and median-ISDAR-LIB) outperformed the single-loudspeaker method (i.e. ISDAR-LIB). Mean-ISDAR-LIB was the better image-source reversion reflector locator, among those tested. Grouping by room size (see Table \ref{tab:simulated_datasets}), we note that every method suffers when the room dimensions become too small. This is due to the fact that, in really small environments, the loudspeakers, which are perfectly omnidirectional for the simulated datasets, can happen to be closer to different reflectors, raising an ambiguity on which reflector is under investigation. ETSAC, the direct locator, is still better under these conditions. On the other hand, organizing the results considering the three different $\overline{\eta}$, when $\overline{\eta}=0.5$ all the methods seem to deteriorate. However, as shown in Table \ref{tab:simulated_datasets}, the smallest room generated has been coincidentally selected to have $\overline{\eta}=0.5$, and there is no clear trend between $\overline{\eta}=0.2$ and $\overline{\eta}=0.8$ under these conditions. The methods are more affected by the room size rather than $\overline{\eta}$. Again, the direct localization ETSAC is the best method under every condition. The $\mu_{\mathrm{RMSE}}$ reported on the bottom of the table, should be read with the related $G_\mathrm{RMSE}$, as the RMSE of the fine error values depends on the amount of gross errors eliminated from the calculation. First, median-ISDAR-LIB has consistent results over all the conditions: although it produces gross errors with more datasets than mean-ISDAR-LIB, if the setup gives fine errors it is more robust on identifying outliers over the estimated image sources. Compared to the image-source reversion method with lowest $G_\mathrm{RMSE}$, ETSAC's fine error is better. There is also a tendency for higher $\overline{\eta}$s to produce higher fine errors with every method.  

For the second set of simulations, observing the $G_\mathrm{RMSE}$ reported in the top part of Table \ref{tab:simulated_SNR}, the only two methods that are not strongly affected by lower DNRs are mean-ISDAR-LIB and ETSAC. ETSAC is, in general the best method here tested, however, it faces small issues with $\mathrm{DNR}=30$\,dB. Here, mean-ISDAR-LIB has comparable performance, showing a high robustness over DNR variations. Nevertheless, looking at the fine errors on the right side of the table, a general trend of improving performance with increasing DNR can be noted. The only one that does not follow that trend is ETSAC. However, it has the lowest $G_\mathrm{RMSE}$ for $\mathrm{DNR}=40$\,dB and $\mathrm{DNR}=50$\,dB, which includes more samples in its $\mu_{\mathrm{RMSE}}$ calculation. Compared to mean-ISDAR-LIB, ETSAC has lower $\mu_{\mathrm{RMSE}}$, showing ETSAC to be the best method tested in these simulations. Given that the first reflection can be 10-20\,dB down from the direct sound \cite{HowAng2009}, reflector estimation may be expected to degrade at DNRs below 20\,dB.    

\begin{table}[!t]

\caption{Reflector localization gross error $\bm{G_{\mathrm{RMSE}}}$, averaged RMSE $\bm{\mu_{\mathrm{RMSE}}}$, overall for the simulated dataset, varying the DNR (30\,dB, 40\,dB and 50\,dB). The methods, A, B, C and D, are as in Table \ref{tab:simulated_reflector_localization}.}
\label{tab:simulated_SNR}
\centering

\begin{tabular}{|c|c|c|c||c|c|c|}
\cline{2-7}
\multicolumn{1}{c|}{}& \multicolumn{3}{c||}{$\bm{G_{\mathrm{RMSE}}}$ $\bm{(\%)}$} & \multicolumn{3}{c|}{$\bm{\mu_{\mathrm{RMSE}}}$ $\mathbf{\pm}$ $\bm{\mathrm{CI}_{\mathrm{RMSE}}}$ \textbf{(mm)}}   \\ 
\cline{2-7}
\multicolumn{1}{c|}{}& \textbf{30\,dB} & \textbf{40\,dB} & \textbf{50\,dB} & \textbf{30\,dB} & \textbf{40\,dB} & \textbf{50\,dB} \\
\hline
\textbf{A} & 11.8 & 9.4 & 9.3 & $41\pm 2$ & $28\pm 2$ & $30\pm 2$ \\
\hline
\textbf{B} & 19.3 & 5.1 & 5.7 & $45\pm 2$ & $33\pm 1$ & $33\pm 1$ \\
\hline
\textbf{C} & \textbf{3.3} & 3.9 & 3.6 & $128\pm 6$ & $107\pm 6$ & $109\pm 6$ \\
\hline
\textbf{D} & 3.7 & \textbf{2.1} & \textbf{2.2} & $65\pm 1$ & $80\pm 1$ & $80\pm 1$ \\
\hline

\end{tabular}
\end{table}

\subsection{Comparative Evaluation with Recorded Data}
\label{subsec:evaluation}
In this paper, only a subset of the 3D methods presented in Table \ref{tab:3d_models} are evaluated. The other methods are based on assumptions which tend to be too restrictive: in \cite{KusdeVHulGis2004} it was assumed that a uniform linear array of microphones was placed parallel to the reflector; in \cite{ZamanAnnRab2014} they assumed only two surfaces in the room to be reflective; and the method in \cite{RibFloBaZha2012} required large datasets. An evaluation of the direct localization baseline method \cite{NasAntSarTub2011} has been already presented in our recent work \cite{RemJackWangChamb2015}, and therefore omitted here. Consequently, the methods compared here are Tervo \emph{et al.} \cite{TervoTossa2012}, Dokmani\'{c} \emph{et al.} \cite{DokParWalLuVet2013}, and our proposed ones. As described in Section \ref{subsec:pre-processing}, C-DYPSA and DSB were applied as pre-processors, to provide coherent input to every method. 

The comparative evaluation was performed in three main parts. First, the three image-source reversion methods, based on a single loudspeaker were evaluated (i.e. Tervo \emph{et al.} \cite{TervoTossa2012}, Dokmani\'{c} \emph{et al.} \cite{DokParWalLuVet2013}, and the proposed ISDAR-LIB). Keeping in common the reflector localization part of these methods (i.e. LIB), their image source locator algorithms were assessed. Second, the ISDAR-LIB variants, together with the single loudspeaker version, and the direct localization method, were compared, to determine which conceptual approach is the best to perform the reflector location. The third experiment observes, given the plane generated by the best method, whether the corresponding image source is closer to the groundtruth, compared to the one localized with the three image locator algorithms.

\subsubsection{Image Source Localization}
\label{subsubsec:image_experiments}
The single loudspeaker image-source reversion methods are compared using their gross $G_{\epsilon}$ and fine $\mu_{\epsilon}$ errors. The TOA estimator C-DYPSA is used as a pre-processor by every tested method. Therefore, although the performance of the ML algorithm and the multilateration are lower than our proposed methods, this difference cannot be attributed to a large error variance produced by C-DYPSA. As in Table \ref{tab:reproducibility_params}, we used $x=10^4$ sample points for the ML algorithm, 10 times more than in \cite{TerPatLok2012}. The results are reported within the first three rows of Table \ref{tab:image_localization_ETSAC}, showing that ISDAR performs much better than the two baselines, benefitting from two acoustic parameters (i.e. TOA and DOA), rather than only TOA. ``VML'' appears as a problematic dataset for every algorithm tested, due to its high reverberance at the middle-high frequencies. Furthermore, although the ML algorithm \cite{TervoTossa2012} and ISDAR provide some fine errors as output, the multilateration fails~\cite{DokParWalLuVet2013} with $G_\epsilon=100\%$. In Table \ref{tab:image_localization_ETSAC}, the weighted average (W-AVG) error over all the rooms is also reported, calculated taking into account the amount of fine errors $\epsilon_j$ provided by each dataset.

\begin{table}[!t]

\caption{Image source localization gross error $\bm{G_\epsilon}$, averaged error $\bm{\mu_\epsilon}$, and confidence interval $\bm{\mathrm{CI}_\epsilon}$, related to the four recorded datasets, and their weighted average (W-AVG). The methods \rom{1}, \rom{2}, \rom{3}, and \rom{4} are Maximum Likelihood \cite{TervoTossa2012}, Multilateration \cite{DokParWalLuVet2013}, ISDAR, and Mirrored ETSAC, respectively. }
\label{tab:image_localization_ETSAC}
\centering

\begin{tabular}{|c|c|c|c|c|c|}
\cline{2-6}
\multicolumn{1}{c|}{\textbf{$\bm{G_\epsilon}$} $\bm{(\%)}$}& \textbf{ABooth} & \textbf{Vislab} & \textbf{VML} & \textbf{Studio1} & \textbf{AVG} \\
\hline
\textbf{\rom{1}} & 67.7 & 70.3 & 89.0 & 66.0 & $73.3\pm 9.0$ \\
\hline
\textbf{\rom{2}} & 18.8 & 25.8 & 100.0 & 5.8 & $37.6\pm 36.0$ \\
\hline
\textbf{\rom{3}} & \textbf{0.0} & \textbf{0.0} & 68.2 & \textbf{0.0} & $17.1\pm 28.9$ \\
\hline
\textbf{\rom{4}} & \textbf{0.0} & \textbf{0.0} & \textbf{50.0} & \textbf{0.0} & $\bm{12.5\pm 21.2}$ \\
\hline
\hline
\multicolumn{1}{c|}{$\bm{\mu_\epsilon}$ \textbf{(mm)}}& \textbf{ABooth} & \textbf{Vislab} & \textbf{VML} & \textbf{Studio1} & \textbf{W-AVG} \\
\hline
\textbf{\rom{1}} & $323$ & $328$ & $\mathbf{342}$ & $331$ & $334\pm 6$ \\
\hline
\textbf{\rom{2}} & $265$ & $263$ & -- & $296$ & $267\pm 10$ \\
\hline
\textbf{\rom{3}} & $208$ & $239$ & $352$ & $232$ & $245\pm 4$ \\
\hline
\textbf{\rom{4}} & $\mathbf{82}$ & $\mathbf{163}$ & $438$ & $\mathbf{100}$ & $\mathbf{220\pm 8}$ \\
\hline
\end{tabular}
\end{table}

\subsubsection{Reflector Localization}
Having identified the novel ISDAR-LIB as the best single-loudspeaker image-source reversion method, it is then compared with its two novel variants (i.e. mean-ISDAR-LIB and median-ISDAR-LIB), that utilize the information from multiple loudspeakers, and our direct localization method ETSAC. The results are reported in Table \ref{tab:reflector_localization}, where the $\mu_{\mathrm{RMSE}}$ values are calculated following Equation~(\ref{eq:AVG_RMSE_reflectors}). For every dataset and every method $G_{\mathrm{RMSE}}=0\%$.  

The results show that ETSAC performs much better than the other methods. This indicates that the better approach to localize reflectors, for these compact microphone array RIRs, is the direct localization rather than the image-source reversion. On the other hand, it is not possible to distinguish which method is the best among the image-source reversion methods. Every dataset provides different results. 
However, observing the $\mu_{\mathrm{RMSE}}$ averaged over all the datasets, mean-ISDAR-LIB performs best. For the ``Vislab'', the introduction of multiple loudspeakers did not have a noticeable effect on the image-source reversion method results (even though it reduced the $\mathrm{CI}_{\mathrm{RMSE}}$). This is due to the fact that LIB performs similarly with every loudspeaker in this room. ``Studio1'' is a dataset including four loudspeakers. Due to this small number, methods that use multiple loudspeaker information do not obtain improvement. With the ``AudioBooth'' there are problems in LIB with the correct identification of the normal vectors $\textbf{v}_{j}$, however, the midpoints $\textbf{M}_{j}$ are finely localized. The median-ISDAR-LIB method, which exploits the median of $\textbf{M}_{j}$, gave lower performance than the others, since it is not robust to fine errors. Finally, ``VML'' is again the most problematic dataset. However, even with this dataset ETSAC has better performance. Nevertheless, for one loudspeaker, our implementation of ISDAR-LIB has a run time of 11\,ms, whereas ETSAC requires 2.1\,s, making ISDAR-LIB appealing for fast processing purposes.     

In addition to mean- and median-ISDAR-LIB, two other ISDAR-LIB variants were tested, fitting a plane to the $L$ midpoints. The first used least squares, the second used RANSAC. Although they improved over ISDAR-LIB, their performance was lower than mean- and median-ISDAR-LIB. 

\begin{table}[!t]

\caption{Reflector localization averaged RMSE $\bm{\mu_{\mathrm{RMSE}}}$, and confidence interval $\bm{\mathrm{CI}_\mathrm{RMSE}}$, related to the four datasets, and their average (AVG). The methods A, B, C, and D, are as explained in Table \ref{tab:simulated_reflector_localization}.}
\label{tab:reflector_localization}
\centering
\resizebox{\columnwidth}{!}{

\begin{tabular}{|c|c|c|c|c|c|c|c|}
\cline{2-6}
\multicolumn{1}{c|}{$\bm{\mu_{\mathrm{RMSE}}}$ \textbf{(mm)}}& \textbf{ABooth} & \textbf{Vislab} & \textbf{VML} & \textbf{Studio1} & \textbf{AVG} \\
\hline
\textbf{A} & $86$ & $47$ & $148$ & $46$ & $102\pm 20$ \\
\hline
\textbf{B} & $92$ & $70$ & $120$ & $54$ & $96\pm 10$ \\
\hline
\textbf{C} & $56$ & $59$ & $127$ & $49$ & $90\pm 12$ \\
\hline
\textbf{D} & $\mathbf{21}$ & $\mathbf{30}$ & $\mathbf{82}$ & $\mathbf{17}$ & $\mathbf{52\pm 2}$ \\
\hline

\end{tabular}
}
\end{table}

\subsubsection{Image Source Localization, a Cross-Check}
\begin{figure}[tp] 

\centering
\subfloat[AudioBooth. Modified from \cite{RemJacColFra2015}.]{%
\includegraphics[trim={435px 195px 400px 160px},clip,width=0.39\textwidth]{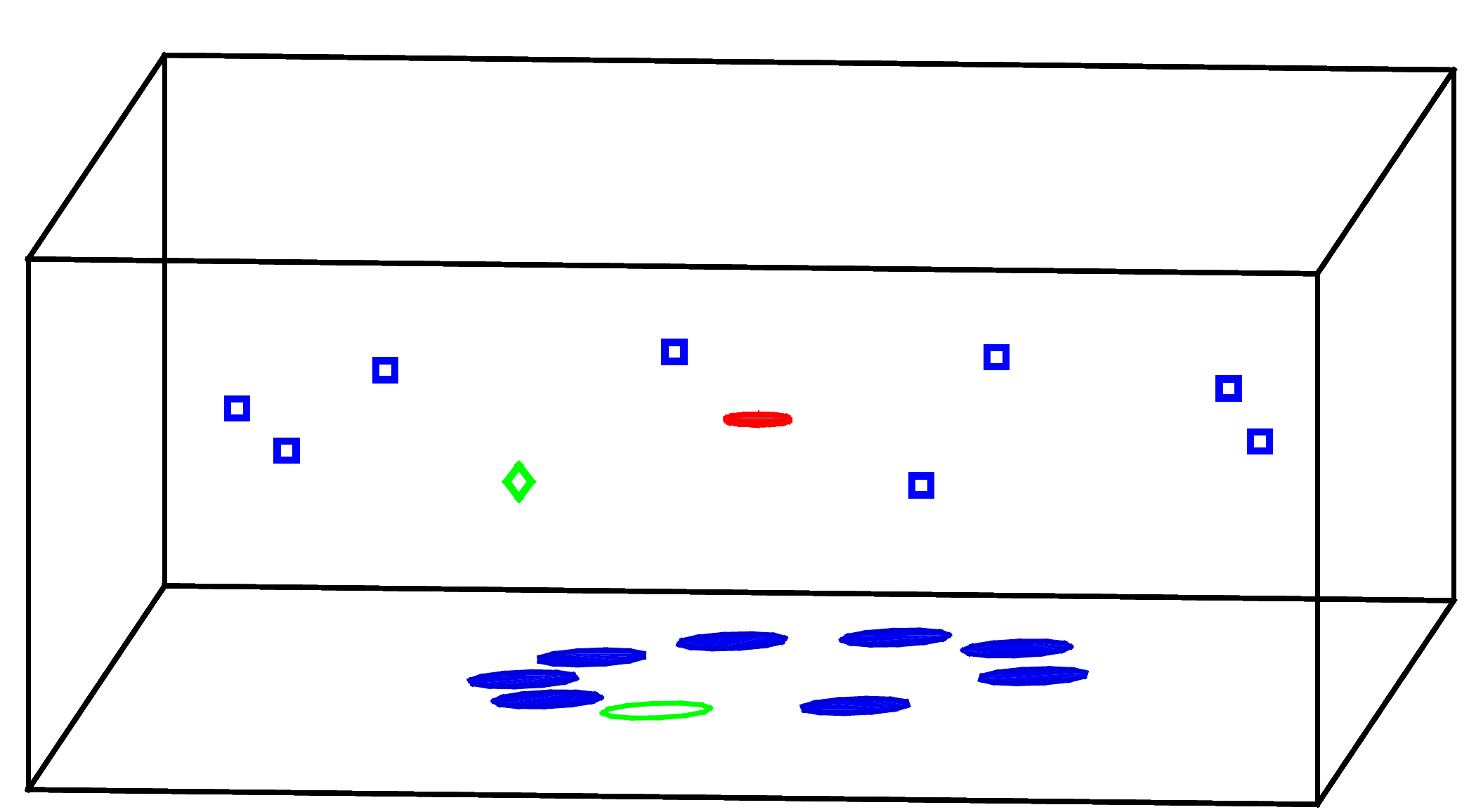}%
\label{fig:ebrief_like_AudioBooth}%
}
\hspace{-30pt}
\subfloat[VML.]{%
\includegraphics[trim={500px 165px 450px 200px},clip,width=.34\textwidth]{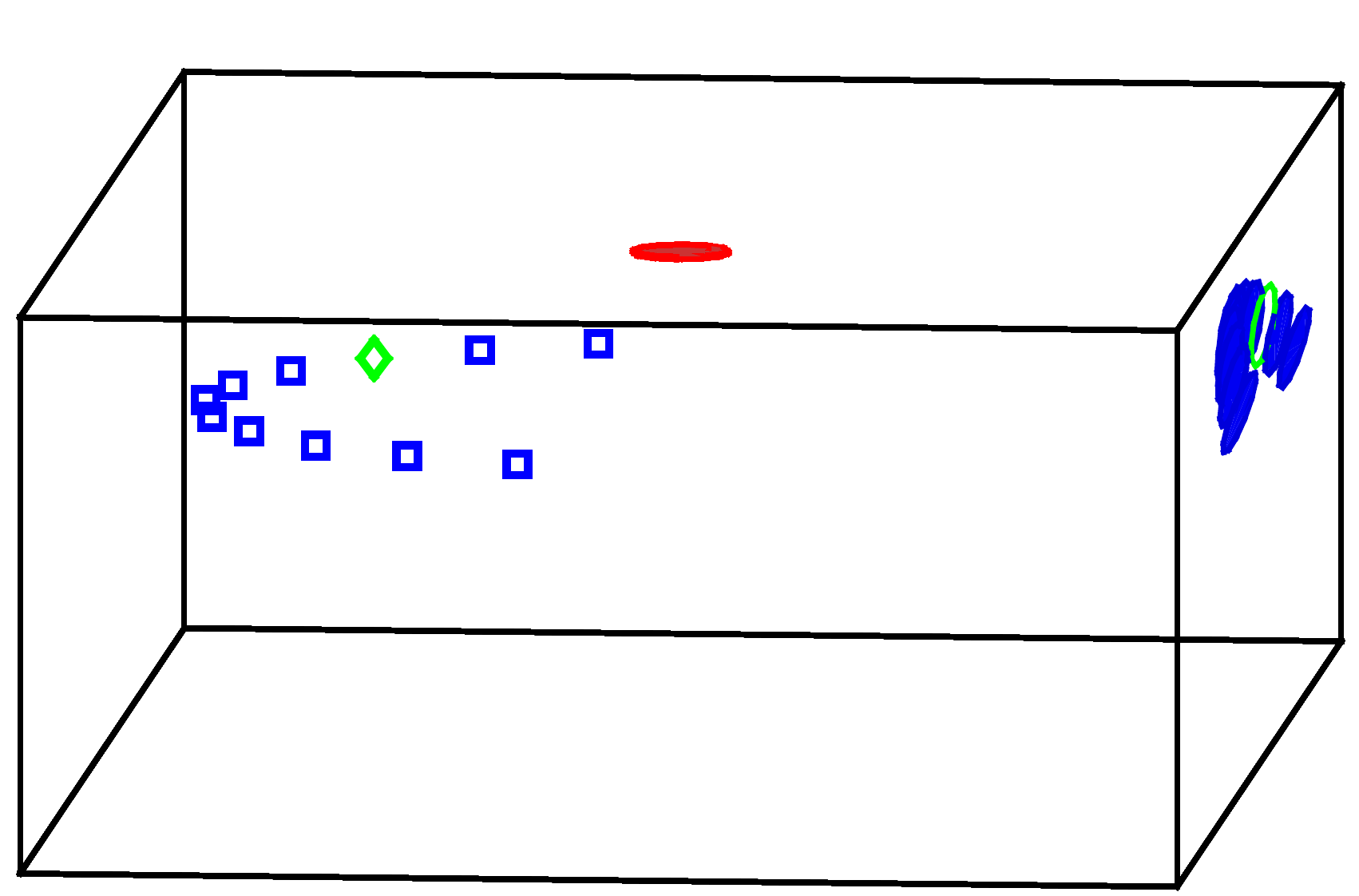}%
\label{fig:ebrief_like_VML}%
}
\caption{Overview of the reflection positions estimated using ETSAC (blue circles) for two recorded datasets. The green circle represents the reflection position related to the loudspeaker left out during the LOLO estimation (green diamond). The loudspeaker positions (blue squares) and the microphone array (red circles) are represented inside the room geometry groundtruths (boxes).}
\label{fig:visualization_ebrief_like}
\end{figure}

To evaluate the ETSAC performance directly together with \cite{TervoTossa2012} and \cite{DokParWalLuVet2013}, images were calculated from the estimated plane. In particular, having all the $\textbf{p}_{l}$ from running ETSAC, where $l$ is the index of the loudspeaker combination, and the $L$ loudspeaker positions $\textbf{B}_{0,j}$, the $L$ images $\textbf{B}_{j}$ were localized. Then, exploiting Equation~(\ref{eq:mid-point}), the midpoints $\textbf{M}_{j}$ were obtained. This method is named Mirrored ETSAC. Figure \ref{fig:visualization_ebrief_like} shows circles to mark reflection positions in the plane of the estimated reflector with a shoebox outline of each room. 

The image localization errors for Mirrored ETSAC, calculated as before by Equation~(\ref{eq:error_images}), are reported in the last row of Table \ref{tab:image_localization_ETSAC}. The fine error results indicate that the images generated via the ETSAC-estimated reflector are consistently more accurate than those from the other methods in AudioBooth, Vislab and Studio1. In VML, we observe an increment on the level of fine errors since all the methods find this dataset challenging, as seen in the high levels of gross error. The key result here therefore is the reduction in gross error rate, from over two thirds down to one half using Mirrored ETSAC. The gross error reduction by Mirrored ETSAC however comes with sacrifice in its fine error score in VML. Nevertheless, the W-AVG shows an overall improvement in the performance. In addition, comparing only those cases that both ISDAR-LIB and Mirrored ETSAC successfully resolved (i.e., discarding any cases of gross error), the fine error average are 248$\pm$5 and 160$\pm$7 respectively, which confirms the superior performance of Mirrored ETSAC. In spite of this, with a gross error rate of 12.5\% across all four datasets, we observe an average error of 22\,cm in the image source location, whereas it is 5\,cm in the reflector location. It is also interesting to note that the scale of these localization errors is comparable to the limits of human perception \cite{MakMid1990}.

\subsubsection{Computational Complexity}
To assess the computational complexity of image-source reversion and direct localization methods, a rough calculation of the number of linear and non-linear operations is reported, considering ISDAR-LIB and ETSAC. ISDAR-LIB needs a total of 21 linear operations and 8 non-linear ones to find the reflector. On the other hand, ETSAC employs 83 linear and 9 non-linear operations to generate each of the $N=M\cdot L$ ellipsoids, plus 93 linear and 2 non-linear operations for each $p$-th plane generated. In addition, ETSAC uses, in the reflector search step, a sampling method based on RANSAC, which tests $P=10^4$ planes before finding the best one. On the other hand, ISDAR-LIB, once it localizes the image source, it estimates the position of the related plane once. As a result, ISDAR-LIB had a run time approximately 200 times faster than ETSAC (i.e. the run times are 0.011\,s for ISDAR-LIB, and 2.123\,s for ETSAC).

\section{Conclusion}
\label{sec:concl}
We presented four novel reflector localization methods: three image-source reversion (ISDAR-LIB, mean-ISDAR-LIB, median-ISDAR-LIB), and a direct localization (ETSAC). To automatically extract TOAs from multichannel RIRs, the novel C-DYPSA was also introduced. The proposed methods were compared with baselines, to discover the best approach for reflector localization given compact microphone array RIRs. 

Simulations of recording conditions, with background noise and microphone positions displacements, were used to test the methods by varying the room size, absorption coefficient and DNR. Results showed that ETSAC performed better than the other methods tested, in every condition. All methods were affected by gross errors for small environments, fine errors increased with the increasing in absorption coefficient. Furthermore, mean-ISDAR-LIB and ETSAC were robust to low DNR conditions.
Experiments with recorded RIRs were divided into three main tasks. Firstly, the image localization algorithms proposed in \cite{TervoTossa2012} and \cite{DokParWalLuVet2013} were compared to ISDAR. Results show that the novel ISDAR provided the best performance. The second part of the experiments compared ISDAR-LIB, with mean-ISDAR-LIB, median-ISDAR-LIB, which are novel image-source reversion methods exploiting multiple loudspeakers, together with our direct localization method ETSAC. Results show that ETSAC localized the reflector with an average 5\,cm RMSE, i.e., 42\% lower than the best alternative method, here tested. In the last experiment, the reflectors estimated through ETSAC were converted into their corresponding image sources, and compared with the image locators in \cite{TervoTossa2012} and \cite{DokParWalLuVet2013}. This showed the percentage of gross errors dropping drastically from 38\% (multilateration \cite{DokParWalLuVet2013}) to 13\% (ETSAC). To sum up, these experiments showed that the direct localization gave better reflector localization accuracy than image-source reversion across the evaluated RIR datasets. Results also showed that images then located by mirroring sources in the reflector also benefited from the improved reflector estimation. However as ISDAR-LIB ran 200 times faster than ETSAC, it has an advantage for fast processing applications, as well as single-source measurements, which may be useful in tracking. 

Further improvements may in the future be found by exploring alternative microphone array arrangements over a large set of rooms, optimal beamformer designs for DOA estimation, and robust methods for multiple loudspeaker ISDAR-LIB post-processing.

\section*{Acknowledgment}
The authors would like to thank Dr.~Sakari~Tervo of Aalto University, who has kindly shared the LIB code. We wish also to thank the associate editor and the anonymous reviewers for their contributions to improving this paper.

\ifCLASSOPTIONcaptionsoff
  \newpage
\fi

\bibliographystyle{IEEEbib}

\begin{IEEEbiography}[{\includegraphics[height=1.25in,clip,keepaspectratio]{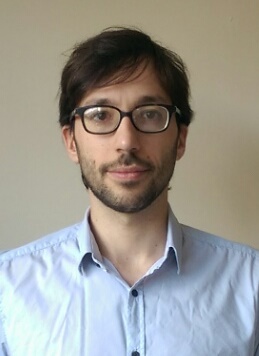}}]{Luca Remaggi}
is currently pursuing the PhD degree at the Centre for Vision, Speech and Signal Processing, University of Surrey, UK. He is investigating the multipath propagation of sound, with focus on various applications, such as spatial audio and blind source separation. Previously, he received the B.Sc. and M.E. degrees in Electronic Engineering from Universit\`{a} Politecnica delle Marche, Italy, in 2009 and 2012, respectively. During his M.E., he spent some months on an internship, at the Department of Signal Processing and Acoustics, Aalto University, Finland, where he focused on the sound synthesis of musical instruments. He also worked as Researcher at Loccioni Group, Italy, between 2012 and 2013.
\end{IEEEbiography}

\begin{IEEEbiography}[{\includegraphics[height=1.25in,clip,keepaspectratio]{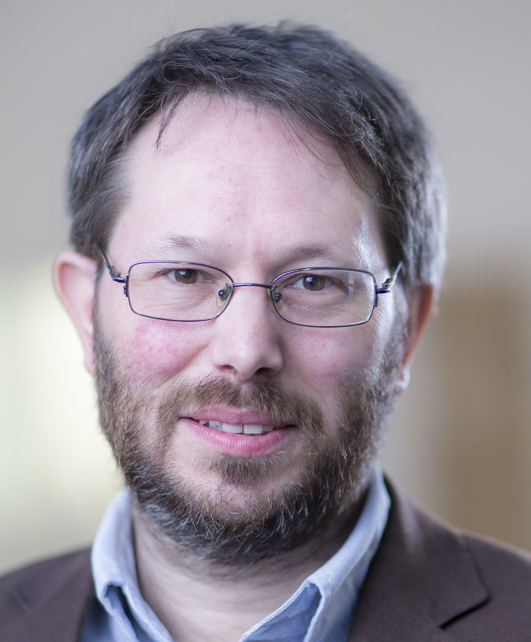}}]{Philip Jackson}
is Senior Lecturer in machine audition at the Centre for Vision, Speech \& Signal Processing (University of Surrey) with MA in Engineering (Cambridge University) and PhD in Electronic Engineering (University of Southampton), all in the UK. His broad interests in acoustical signal processing have yielded research contributions in speech production, processing and recognition, audio-visual machine learning, blind source separation, articulatory modeling, visual speech synthesis, spatial audio reproduction and quality evaluation, and sound field control [https://scholar.google.com/citations?hl=en\&user=n4fuxwQAAAAJ]. He currently leads the object-based capture and production research stream in the S3A project on future spatial audio.

\end{IEEEbiography}
\vspace{-20pt}
%
\begin{IEEEbiography}[{\includegraphics[height=1.25in,clip,keepaspectratio]{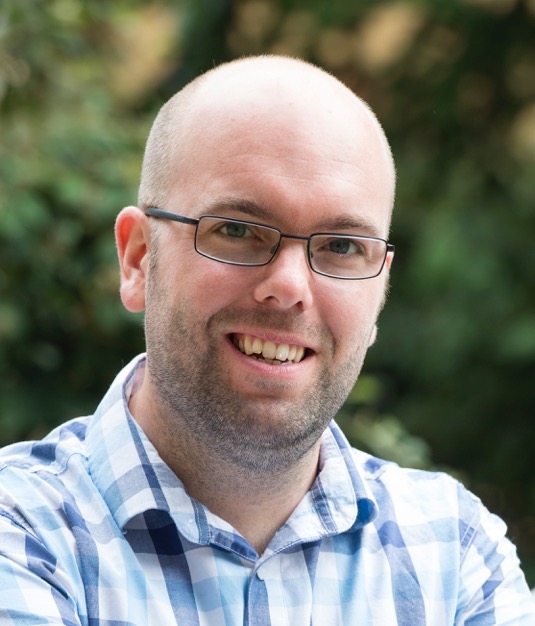}}]{Philip Coleman}
joined the Centre for Vision, Speech and Signal Processing, University of Surrey, UK, in 2010, earning his PhD in 2014 on the topic of personal sound zones. He is currently working in the centre as a research fellow on the project S3A: Future spatial audio for an immersive listening experience at home, with a focus on recording and editing object-based content. His research interests include sound field control, loudspeaker and microphone array processing, and spatial audio. Previously, he received the BEng degree in Electronic Engineering with Music Technology Systems in 2008 from the University of York, UK, and MSc with distinction in Multimedia Signal Processing and Communication from the University of Surrey, UK, in 2010.
\end{IEEEbiography}
\vspace{-20pt}
\begin{IEEEbiography}[{\includegraphics[height=1.25in,clip,keepaspectratio]{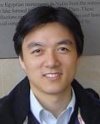}}]{Wenwu Wang}
(M'02–-SM'11) was born in Anhui, China. He received the B.Sc. degree in automatic control in 1997, the M.E. degree in control science and control engineering in 2000, and the Ph.D. degree in navigation guidance and control in 2002, all from Harbin Engineering University, Harbin, China. 
He then joined King's College, London, U.K., in May 2002, as a postdoctoral research associate and transferred to Cardiff University, Cardiff, U.K., in January 2004, where he worked in the area of blind signal processing. In May 2005, he joined the Tao Group Ltd. (now Antix Labs Ltd.), Reading, U.K., as a DSP engineer working on algorithm design and implementation for real-time and embedded audio and visual systems. In September 2006, he joined Creative Labs, Ltd., Egham, U.K., as an R\&D engineer, working on 3D spatial audio for mobile devices. Since May 2007, he has been with the Centre for Vision Speech and Signal Processing, University of Surrey, Guildford, U.K., where he is currently a Reader in Signal Processing, and a Co-Director of the Machine Audition Lab. He is a member of the Ministry of Defence (MoD) University Defence Research Collaboration (UDRC) in Signal Processing (since 2009), a member of the BBC Audio Research Partnership (since 2011), an associate member of Surrey Centre for Cyber Security (since 2014), and a member of the MRC/EPSRC Microphone Network (since 2015). During spring 2008, he has been a visiting scholar at the Perception and Neurodynamics Lab and the Center for Cognitive Science, The Ohio State University. 
His current research interests include blind signal processing, sparse signal processing, audio-visual signal processing, machine learning and perception, machine audition (listening), and statistical anomaly detection. He has (co)-authored over 150 publications in these areas, including two books \textit{Machine Audition: Principles, Algorithms and Systems} (IGI Global, 2010) and \textit{Blind Source Separation: Advances in Theory, Algorithms and Applications} (Springer, 2014). He is currently an Associate Editor for IEEE Transactions on Signal Processing. He is also Publication Co-Chair of ICASSP 2019 (to be held in Brighton, UK). He was a Tutorial Speaker on ICASSP 2013, UDRC Summer School 2014, 2015 and 2016, and SpaRTan/MacSeNet Spring School 2016.

\end{IEEEbiography}

\end{document}